\documentclass[fleq,usenatbib]{mnras}
\usepackage{newtxtext,newtxmath}
\usepackage[T1]{fontenc}

\DeclareRobustCommand{\VAN}[3]{#2}
\let\VANthebibliography\thebibliography
\def\thebibliography{\DeclareRobustCommand{\VAN}[3]{##3}\VANthebibliography}

\usepackage{graphicx}
\usepackage{dcolumn}
\usepackage{comment}
\usepackage{bm}
\usepackage{hyperref}
\usepackage{float}
\usepackage{amsmath}

\usepackage{mathtools}

\newcommand{\md}{\mathrm{d}}

\def\p{\partial}
\newcommand{\pder}[2]{\frac{\partial#1}{\partial#2}}

\newcommand{\kms}{\mathrm{km}\, \mathrm{s}^{-1}}

\newcommand{\esq}{\varepsilon}

\newcommand{\nn}{\nonumber}
\def\p{\partial}

\newcommand{\Msun}{M_\odot}

\newcommand{\C}{\overline{C}}

\newcommand{\meanDelta}[1]{\langle{\Delta #1}\rangle}
\newcommand{\meansquareDelta}[1]{\langle{(\Delta #1)^2}\rangle}
\newcommand{\orbitmeanDelta}[1]{\overline{\langle \Delta #1\rangle}}
\newcommand{\orbitmeansquareDelta}[1]{\overline{\langle{(\Delta #1)^2}\rangle}}
\newcommand{\tion}{t_\mathrm{dis}}

\usepackage{accents}

\title[Eccentricity of wide binaries - II. Stellar encounters]{Eccentricity dynamics of wide binaries - II. The effect of stellar encounters and constraints on formation channels}

\author[C. Hamilton \& S. Modak]{
Chris Hamilton$^{1}$\thanks{E-mail: chamilton@ias.edu} and 
Shaunak Modak$^{2}$
\\
$^{1}$ Institute for Advanced Study, Einstein Drive, Princeton, NJ 08540, USA
\\
$^{2}$ Department of Astrophysical Sciences, Princeton University, 4 Ivy Lane, Princeton, NJ 08544, USA\\
}

\pubyear{2023}

\begin{document}
\graphicspath{{./}{figures/}}
\label{firstpage}
\pagerange{\pageref{firstpage}--\pageref{lastpage}}
\maketitle 


\begin{abstract}
GAIA wide stellar binaries {(separations $\sim 10^3-10^{4.5}$ AU) are observed to have} a superthermal eccentricity distribution function (DF), well-fit by $P(e) \propto e^\alpha$ with $\alpha \sim 1.2$. In {\cite{MH23}}, we proved that this DF cannot have been produced by Galactic tidal torques starting from any realistic DF that was not already superthermal. Here, we consider the other major dynamical effect on wide binaries: encounters with passing stars. We derive and solve the Fokker-Planck equation governing the evolution of binaries in semimajor axis and eccentricity under many weak, impulsive, penetrative stellar encounters. We show analytically that these encounters drive the eccentricity DF towards thermal on the same timescale as they drive the semimajor axes $a$ towards disruption, $\tion \sim 4\,\mathrm{Gyr}\,(a/10^4\,\mathrm{AU})^{-1}$. We conclude that the observed superthermal DF must derive from an even more superthermal (i.e. higher $\alpha$) birth distribution. This requirement places strong constraints on the dominant binary formation channels. A testable prediction of our theory is that $\alpha$ should be a monotonically decreasing function of binary age.
\end{abstract}




\section{Introduction}
\label{sec:Introduction}


GAIA has allowed astronomers to measure for the first time the eccentricity distributions of wide stellar binaries in the Solar neighborhood. {For binary separations in the range ($10^3, 10^{4.5}$AU), the
eccentricity distribution function (DF) is found to be robustly \textit{superthermal}, well-fit by a power law $P(e) = (1+\alpha)e^{\alpha}$, with $\alpha \sim 1.2$} \citep{Tokovinin2020-go,Hwang21}\footnote{{In \cite{MH23} we quoted $\alpha \sim 1.3$, but a small systematic bias in the measurements means that the true $\alpha$ value is likely slightly smaller than this \citep{Hwang21}. Also,  $\alpha$ is probably weakly separation-dependent --- see  \S\ref{sec:Astrophysical_Implications}.}}. {Superthermal DFs contain more eccentric binaries than the so-called} \textit{thermal} distribution, $P(e) = 2e$, which one expects if binaries are spread out uniformly in phase space at fixed $a$. The origin of the observed superthermal DF is unknown. In particular, it was previously unclear whether superthermality is the result of dynamical evolution, or is a symptom of the wide binary formation process(es) \citep{Hwang21}.

Dynamical perturbations to (bound) wide binaries essentially fall into two categories: (i) the coherent torque of the Galactic tide, and (ii) encounters with passing stars, molecular clouds, and so on. Most previous theoretical work looking at the impact of effects (i) and (ii) upon wide binaries \citep{Weinberg1987, jiang2010evolution} focused on the evolution of semimajor axes rather than eccentricity --- in fact, they assumed a fixed thermal eccentricity distribution throughout. In contrast, \cite{Hamilton22} started from an arbitrary initial eccentricity DF, and calculated the steady-state DF that one would expect if binaries were perturbed by Galactic tides alone (effect (i)). He found that Galactic tides cannot themselves \textit{produce} a superthermal DF from any DF that was not initially already superthermal. More recently, we have confirmed and significantly extended this result (\citealt{MH23}, hereafter Paper I). Briefly, the upshot of \cite{Hamilton22} and Paper I is that for initially isotropically-oriented, Solar-mass binaries, an initial power law eccentricity DF with (effective) index $\alpha_\mathrm{i}$ is transformed under Galactic tidal torques alone into another power law with index $\approx (1+\alpha_\mathrm{i})/2$, on a timescale $ \sim 4\,\mathrm{Gyr}\,(a/10^4\,\mathrm{AU})^{-3/2}$. It follows that the Galactic tide cannot transform any sensible `subthermal' DF ($\alpha_\mathrm{i} < 1$) into the observed superthermal one.

The primary aim of the present paper is to investigate the other major dynamical effect upon wide binaries, namely stellar encounters. A priori, it is not obvious what the effect of these encounters upon the eccentricity DF should be. For instance, if one fixed the binaries' semimajor axes $a$ (or equivalently their binding energies $E$), and then allowed many weak encounters to destroy any nonuniformity in the remaining degrees of freedom, the eccentricity DF at each $a$ would inevitably tend towards thermal. But the binaries we are considering are so wide that stellar encounters are \textit{impulsive} (meaning each encounter occurs on a timescale short compared to the binary orbital period), and impulsive kicks \textit{do} change $a$; in fact, kicks to $a$ and $e$ are correlated. What is more, impulsive kicks produce a systematic drift towards higher $a$ (less negative $E$), so soft binaries become softer \citep{heggie1975binary} until they are eventually disrupted by the Galactic tide at around $a\sim 2\times 10^5$AU \citep{jiang2010evolution}. This flux towards larger $a$ and eventual disruption means that in the absence of a constant source of new binaries, no steady state distribution function is possible. Nevertheless, by deriving and solving a Fokker-Planck equation we are able to \textit{prove} {--- subject to a few minor caveats ---} that weak, impulsive, penetrative, isotropic stellar encounters do indeed thermalize the eccentricity distribution of wide binaries at each $a$ while \textit{simultaneously} driving individual binaries towards disruption. We also demonstrate that these two processes occur on essentially the same timescale. This is true regardless of the choice of initial DF.

As for wide binary formation processes, the dominant channels are still unknown, but the possibilities include the dissolution of young stellar clusters \citep{kouwenhoven2010formation,cournoyer2021implementing}, the unfolding of hierarchical triples \citep{reipurth2012formation}, the pairing up of stars in tidal streams \citep{penarrubia2021creation}, dynamical capture processes \citep{rozner2023born,ginat2024three,atallah2024binary}, and turbulent fragmentation \citep{Bate2014, xu2023wide} --- see \S\ref{sec:Discussion} for more, and \cite{Hwang21} for a review.  While we remain agnostic about the formation processes throughout our dynamical analyses, the conclusions we draw (combined with those of Paper I) imply that whatever the formation mechanism, the population of wide binaries in the past must have been significantly more superthermal (higher $\alpha$) than is observed today. This requirement places a strong constraint on any proposed formation channel. The secondary aim of this paper is therefore to use this constraint to rule out various proposed channels, and to devise a testable prediction that must be satisfied if our theory is correct.

The rest of this paper is organized as follows. In \S\ref{sec:approximations} we provide order-of-magnitude estimates of the effect of stellar encounters on wide binaries in order to justify the approximations we use thereafter. Next, in \S\ref{sec:Kinetic_Theory} we derive the Fokker-Planck equation governing the evolution of an ensemble of binaries in $(a,e^2)$ space, investigate some of its properties, and solve it numerically for some simple example initial DFs. Finally, in \S\ref{sec:Discussion} we use our results to constrain wide binary formation channels, make predictions for future measurements, and compare our work to the previous literature. We summarize in \S\ref{sec:summary}.


\section{Timescales and approximations}
\label{sec:approximations}


In this paper we will adopt several simplifying assumptions that allow us to describe binary evolution using a Fokker-Planck equation in semimajor axis and eccentricity. Namely, we assume that encounters with passing stars are approximately:
\begin{itemize}
    \item \textit{Isotropic}. Our wide binaries are affected by encounters from an isotropic distribution of stars moving in straight lines with velocity dispersion $\sigma$ and number density $n$.
    \item \textit{Impulsive}. Each encounter takes place on a timescale much shorter than the binary orbital period.
    \item \textit{Penetrative}. Each binary experiences both penetrative encounters (with impact parameters $b\lesssim a$) and distant encounters ($b \gg a$), but the distant encounters can be ignored.
    \item \textit{Weak}. The relative change to a binary's energy and angular momentum due to a single stellar encounter is small.
\end{itemize}
Moreover, we ignore encounters with molecular clouds and density fluctuations in the interstellar medium, and we ignore the coherent torque due to the Galactic tide (which was the focus of Paper I).

The rest of this section is dedicated to justifying these approximations using order-of-magnitude estimates. A reader who is willing to accept the approximations as reasonable can skip directly to \S\ref{sec:Kinetic_Theory}.


\subsection{Stellar encounters are isotropic and impulsive}
\label{sec:Approximation_Isotropic_Impulsive}


Consider a binary with component masses $m_1$ and $m_2$, semimajor axis $a$, and eccentricity $e$. The specific energy of this binary is 
\begin{align}
    \label{eq:energy}
    E = -\frac{Gm_\mathrm{b}}{2a},
\end{align}
where $m_\mathrm{b} \equiv m_1 + m_2$. Its specific angular momentum vector is $\bm{J} = \bm{r} \times \bm{v}$, where ($\bm{r}, \bm{v}$) are the relative position and velocity vectors of the binary components, and the magnitude of the vector $J = \vert\bm{J}\vert$ may be expressed in terms of its total mass, semimajor axis, and eccentricity as
\begin{equation}  
    \label{eq:J_ae}
    J =  [Gm_\mathrm{b}a(1-e^2)]^{1/2}.
\end{equation}
Throughout this paper, we are interested in binaries with semimajor axes $a\gtrsim 10^3$AU and typical masses $m_\mathrm{b} \sim M_\odot$.

Now let us assume that our binary is surrounded by stars with number density $n$. {The vast majority of the wide binaries in the sample of \cite{Tokovinin2020-go} and \cite{Hwang21} are on disk-like Galactocentric orbits, so we will focus on disk binaries throughout this paper}. Since the scale height of the Galactic disk $\sim 200$\,pc is much larger than any binary under consideration (maximum $a\sim 1$\,pc), we can treat $n$ as a constant. For simplicity we also ignore the effects of secular evolution of the galactic disk and/or the binary's barycentric orbit, both of which would lead to slow evolution of the local $n$.  

We further assume that stars impinge on the binary from random directions and with fixed velocity dispersion $\sigma$ relative to the binary. {These approximations are reasonable since the binaries we are interested in are mostly on disk-like orbits in the Solar neighborhood, where the velocity dispersion $(\sigma \sim 40 \,\mathrm{km/s})$ is much larger than the inner orbital speed of a wide stellar binary ($v_\mathrm{b} \sim 1\,\mathrm{km/s}\,(a/10^3\,\mathrm{AU})^{-1/2}$), and the radial and vertical velocity dispersions are not too different from one another.}\footnote{{There are also some wide binaries in halo-like orbits \citep{hwang2022widethick}. For these a separate analysis would be required, because $n$ is not constant, the relative velocity of perturbers is significantly larger and not necessarily isotropic, etc.}}

Similarly, an encounter between the binary and a passing star will have a typical duration
\begin{equation}
    \label{eq:encounter_time}
    t_\mathrm{enc} \sim \frac{b}{\sigma} \approx 10^3 \, \mathrm{yr}\,\left( \frac{b}{10^4\mathrm{AU}} \right) \left( \frac{\sigma}{40\,\mathrm{km/s}} \right)^{-1},
\end{equation}
where $b$ is the impact parameter of the encounter. Comparing this with the binary's Keplerian orbital period
\begin{equation}
    \label{eq:binary_orbit_time}
    T_\mathrm{b} = 10^6 \, \mathrm{yr} \, \left( \frac{m_\mathrm{b}}{M_\odot} \right)^{-1/2} \left( \frac{a}{10^4 \mathrm{AU}} \right)^{3/2},
\end{equation}
we see that only for extremely distant encounters ($b \gtrsim 10^3 a)$ do we ever have $t_\mathrm{enc} \gtrsim T_\mathrm{b}$; in all other cases the encounters are impulsive, with $t_\mathrm{enc}\ll T_\mathrm{b}$.

An impulsive encounter alters only the binary's relative velocity vector, $\bm{v} \mapsto \bm{v} + \Delta \bm{v}$, without changing its separation vector $\bm{r}$; the corresponding changes to $E$ and $\bm{J}$ in terms of $\Delta \bm{v}$ are given in equation \eqref{eq:DeltaE_DeltaJ}.


\subsection{Stellar encounters are penetrative}
\label{sec:Approximation_Penetrative}


The value of the velocity kick $\Delta \bm{v}$ for a given impulsive stellar encounter depends crucially on whether that encounter's impact parameter, $b$, is small or large compared to the binary semimajor axis $a$. If $b \lesssim a$ then we say the encounter is \textit{penetrative}. Penetrative encounters predominantly affect only one of the stars in the binary, leading to a velocity kick of magnitude
\begin{equation}
    \label{eq:kick_pen}
    \vert \delta \bm{v} \vert_\mathrm{pen}  \approx \frac{2Gm_\mathrm{p}}{ \sigma b},
\end{equation}
where $m_\mathrm{p}$ is the perturber mass. On the other hand, if the impact parameter $b \gg a$ then we say the encounter is \textit{distant}\footnote{Often the word `tidal' is used for these encounters, but we will refrain from using that terminology here to avoid confusion with Galactic tides.}. To lowest order in $a/b$, distant encounters lead to a velocity kick of order
\begin{equation}
    \label{eq:kick_diss}
    \vert \delta \bm{v}\vert_\mathrm{dist} \approx \frac{2Gm_\mathrm{p} a}{ \sigma b^2}.
\end{equation}
To assess the relative importance of penetrative and impulsive kicks for wide binaries, we first estimate the impact parameter of the closest encounter that a binary experiences in time $t$. This is 
\begin{align}
    \label{eq:scatest}
    b_\mathrm{min} & \sim \frac{1}{({\pi n \sigma t})^{1/2}} \nn \\
    & \approx  580 \,\mathrm{AU}\,\left(\frac{n}{0.1 /\mathrm{pc}^{3}}\right)^{-1/2}  \left( \frac{\sigma}{40 \,\mathrm{km/s}}\right)^{-1/2} \left( \frac{t}{10 \, \mathrm{Gyr}}\right)^{-1/2}.
\end{align}
Then binaries with $b_\mathrm{min} \gg a$ will have experienced only distant encounters in their lifetimes, while those with 
$ b_\mathrm{min} \lesssim a $ will have experienced many penetrative encounters as well. The binaries we care about in this paper have $a\gtrsim 10^3\,\mathrm{AU}$ and are at least several Gyr old.  Thus they typically have $b_\mathrm{min} \lesssim a$, though perhaps not by a very large margin for the smallest and youngest binaries.

Let us suppose $b_\mathrm{min} \lesssim a$ and estimate the mean-square velocity kick per unit time, $\langle (\Delta \bm{v})^2\rangle$, felt by a wide binary due to both penetrative and distant encounters. For clarity, we will denote the rough boundary between penetrative and distant encounters by $b = \lambda a$ for some number $\lambda\sim \mathcal{O}(1)$. Then
\begin{equation}
    \label{eq:mean_square_kick_bmin_small}
    \langle (\Delta \bm{v})^2\rangle = \langle (\Delta \bm{v})^2\rangle_\mathrm{pen} + \langle (\Delta \bm{v})^2\rangle_\mathrm{dist},
\end{equation}
where (using equation \eqref{eq:kick_pen}):
\begin{align}
    \label{eq:rms_deltav_pen}
    \langle (\Delta \bm{v})^2\rangle_\mathrm{pen} & \sim  n\sigma \int_{b_\mathrm{min}}^{\lambda a} \md b \, 2\pi b \, \left(\frac{2Gm_\mathrm{p}}{\sigma b}\right)^2 \nn \\
    & = \frac{8\pi G^2 m_\mathrm{p}^2 n }{\sigma }\ln \left(\frac{\lambda a}{b_\mathrm{min}}\right),
\end{align}
and (using equation \eqref{eq:kick_diss}):
\begin{align}
    \label{eq:rms_deltav_dis}
    \langle (\Delta \bm{v})^2\rangle_\mathrm{dist} &\sim  n \sigma \int_{\lambda a}^{b_\mathrm{max}} \md b \, 2\pi b \, \left(\frac{2Gm_\mathrm{p}a}{\sigma b^2}\right)^2 \nn\\
    & \approx  \frac{8\pi G^2 m_\mathrm{p}^2 n }{\sigma } \frac{1}{2\lambda^2},
\end{align}
and in the last line we used $b_\mathrm{max} \gg \lambda a$.
The ratio of the two contributions \eqref{eq:rms_deltav_pen} and \eqref{eq:rms_deltav_dis} is then
\begin{align}
    \frac{\langle (\Delta \bm{v})^2\rangle_\mathrm{dist}}{\langle (\Delta \bm{v})^2\rangle_\mathrm{pen}} &\sim  \frac{1}{2 \lambda^2 \ln (\lambda a/b_\mathrm{min})}.
    \label{eqn:distant_v_penetrative}
\end{align}
Since $\lambda$ is of order (but not less than) unity, and $b_\mathrm{min} \sim 580$\,AU (equation \eqref{eq:scatest}) is much smaller than $a$ for the majority of binaries we are interested in here, the right hand side of equation \eqref{eqn:distant_v_penetrative} is usually small. So, if a binary has experienced penetrative as well as distant encounters, 
then we can reasonably ignore the distant encounters and consider only the effects of the penetrative ones.

In the rare cases where $b_\mathrm{min} \gg a$ (which is only relevant for relatively young binaries with $a\lesssim10^3$AU, see equation \eqref{eq:scatest}), we 
can obviously ignore the penetrative encounters, so that
\begin{equation}
    \label{eq:mean_square_kick_bmin_large}
    \langle (\Delta \bm{v})^2\rangle \sim \frac{8\pi G^2 m_\mathrm{p}^2 n }{\sigma } \frac{a^2}{2b_\mathrm{min}^2}.
\end{equation}
This depends only on the lower limit of the integration $b_\mathrm{min}$, meaning the velocity kick is dominated by the single closest encounter \citep{bahcall1985maximum}.  Nevertheless, it is reassuring that in the intermediate regime  $b_\mathrm{min}\sim a$, the two estimates \eqref{eq:mean_square_kick_bmin_small} and \eqref{eq:mean_square_kick_bmin_large} agree with each other to within an order-unity factor. This gives us confidence that by ignoring the rare cases where 
binaries have \textit{not} experienced penetrative encounters, we are not committing a significant error. Moreover, as we will see, dynamical evolution of binaries in this semimajor axis range is very slow anyway. Hence, we shall ignore distant encounters for the remainder of this paper.


\subsection{Stellar encounters are weak}
\label{sec:Approximation_Weak}


A penetrative, impulsive encounter kicks a binary's velocity vector by an amount $\vert \delta \bm{v}(b)\vert_\mathrm{pen} \sim 2Gm_\mathrm{p}/(b \sigma)$ (equation \eqref{eq:kick_pen}). This kick will be equal to a fraction $\gamma$ of the binary's inner orbital velocity $v_\mathrm{b} \sim (Gm_\mathrm{b}/a)^{1/2}$ for an impact parameter
\begin{align}
    b_\gamma &\sim \frac{2G^{1/2}m_\mathrm{p} a^{1/2}}{\gamma m_\mathrm{b}^{1/2} \sigma} \nn \\
    \label{eq:acat_stars}
    & \approx  \frac{150}{\gamma} \,\mathrm{AU} \,\left( \frac{m_\mathrm{b}}{M_\odot}\right)^{-1/2} \left( \frac{m_\mathrm{p}}{M_\odot}\right) \left( \frac{a}{10^4 \,\mathrm{AU}}\right)^{1/2}  \left( \frac{\sigma}{40 \,\mathrm{km/s}}\right)^{-1}.
\end{align}
{A `catastrophic' kick with impact parameter $b \lesssim b_{\gamma = 1}$ will likely unbind the binary, but equation \eqref{eq:scatest} tells us that such kicks are rare for most binaries of interest, i.e. most encounters are \textit{weak}.}

{Let us be more precise. Following \cite{penarrubia2019stochastic}, we introduce a `fringe' threshold of semimajor axes $a_\mathrm{fringe}$, defined such that a hypothetical binary living for $10$ Gyr at $a=a_\mathrm{fringe}$ would likely experience one strong encounter. By setting $b_{\gamma = 1} = b_\mathrm{min}$, this is
\begin{align}
    \label{eq:fringe}
    a_\mathrm{fringe} \approx 10^5\,\mathrm{AU} & \, \left(\frac{n}{0.1/\mathrm{pc}^3}\right)^{-1}  \left(\frac{\sigma}{40\,\mathrm{km/s}}\right) \nn \\
    & \times \left(\frac{m_b}{\Msun}\right)\left(\frac{m_p}{\Msun}\right)^{-2}\left(\frac{t}{10\,\mathrm{Gyr}}\right)^{-1}.
\end{align}
The encounters experienced by binaries with $a\gtrsim a_\mathrm{fringe}$ cannot safely be considered weak.
In particular, the statistics of such encounters are non-Gaussian \citep{bar2013stellar} and so the evolution of `fringe binaries' is poorly captured by the Fokker-Planck equation we will use in \S\ref{sec:Kinetic_Theory}.  Instead, in this regime one can use turn to something like the Monte-Carlo precription of \cite{penarrubia2019stochastic}, which uses kicks in velocity space rather than energy/angular momentum space\footnote{The advantage being that, for example, 
$(\Delta \mathbf{v})^2 /\vert \mathbf{v}\vert^2 $ does not diverge
as $\vert E\vert \to 0$, whereas $(\Delta E)^2 /\vert E\vert^2 $ does diverge.}, or direct integration like \cite{stegmann2024close}.

{On the other hand, fringe binaries form only a minor subset of all observed wide binaries, mostly because they are disrupted very quickly, on a timescale $\ll 1$ Gyr.  
Indeed, because of low number statistics \cite{Hwang21} only quote eccentricity distributions for binary separations up to $10^{4.5}$ AU, which is below the fringe threshold}\footnote{{Binaries with separations $> 10^{4.5}$AU also appear to be superthermal, but the large error bars make this measurement untrustworthy (H.-C. Hwang, private communication).}}.
{Following their example, in this paper we will restrict all our conclusions to the semimajor axis range $a\in(10^3, 10^{4.5})$ AU.
While we acknowledge that the behavior in the fringe is more complicated --- and probably more violent --- than our model can capture, we do not expect fringe effects to contaminate today's $a\in(10^3, 10^{4.5})$\,AU sample, because (i) once a binary has entered the fringe it almost never returns to smaller $a$, and (ii) since semimajor axis evolution happens more rapidly at large $a$, binaries that are in the range 
$a\in(10^3, 10^{4.5})$ today have spent the vast majority of their lives very far from the fringe. In other words, the great majority of binaries that are not in the fringe today have never been in it in the past.}

{With these caveats}, we can assume that the effect of encounters with stars is weak, i.e. each encounter changes the binary's specific energy and angular momentum by a small amount: $\vert \Delta E \vert \ll \vert E \vert$ and $\vert \Delta \bm{J} \vert \ll \vert \bm{J} \vert$.


\subsection{Neglecting encounters with molecular clouds and Galactic tidal torques}
\label{sec:molecular_clouds}


Isotropic, impulsive, penetrative, weak encounters with stars cause a wide binary with initial semimajor axis $a$ to be disrupted on the timescale (\citealt{BT}, equation (8.59)): 
\begin{align}
    \label{eq:ionize}
    \tion & \approx \frac{0.02\sigma m_\mathrm{b}}{G m_\mathrm{p}^2 n a \ln \Lambda}
     \\
    \label{eq:ionize_numerical}
    & \approx 3.8 \, \mathrm{Gyr}\, \left(\frac{\ln \Lambda}{10} \right)^{-1} \left( \frac{m_\mathrm{b}}{M_\odot}\right) \nn
    \left( \frac{m_\mathrm{p}}{M_\odot}\right)^{-2} \left( \frac{n}{0.1 / \mathrm{pc}^{3}}\right)^{-1} \nn \\ & \,\,\,\,\,\,\,\,\,\,\,\,\,\,\,\,\,\,\,\,\,\,\,\times \left( \frac{\sigma}{40 \,\mathrm{km/s}}\right) \left( \frac{a}{10^4 \mathrm{AU}}\right)^{-1},
\end{align}
where $\Lambda \sim \sigma^2 a/(Gm_\mathrm{p})$ (see equation \eqref{eq:Coulomb} for a precise definition; also \citealt{bahcall1985maximum, Weinberg1987}). Note that this estimate assumes disruption occurs at $E=0$, whereas in reality it occurs once the binary is wide enough to be ripped apart by the Galactic tide ($a\gtrsim 2\times 10^5$\,AU, see \citealt{jiang2010evolution}), {and also ignores the rapid evolution in the fringe (\S\ref{sec:Approximation_Weak}); as a result, the real disruption time is likely slightly shorter than this.}

By contrast, encounters with molecular clouds drive disruption of wide binaries on a timescale (\citealt{BT}, equation (8.62)): 
\begin{align}
    \label{eq:mc}
    t_\mathrm{mc} & \sim  250  \, \mathrm{Gyr}\, \left( \frac{m_\mathrm{b}}{M_\odot}\right) \nn \left( \frac{\rho_\mathrm{mc}}{0.025 \, \Msun/\mathrm{pc}^3}\right)^{-1} \left( \frac{\Sigma_\mathrm{mc}}{300\Msun/\mathrm{pc}^{2}}\right)^{-1} \nn \\
    & \,\,\,\,\,\,\,\,\,\,\,\,\,\,\,\,\,\,\,\,\,\,\,\times   \left( \frac{\sigma}{40 \, \mathrm{km/s}}\right) \left( \frac{a}{10^4\,\mathrm{AU}}\right)^{-3},
\end{align}
where $\rho_\mathrm{mc}$ and $\Sigma_\mathrm{mc}$ are the volumetric and surface mass densities in molecular clouds respectively. This is always much longer than the analogous timescale for stellar encounter-driven evolution (equation \eqref{eq:ionize_numerical}) except for extremely wide, almost-unbound  binaries ($a\gtrsim 10^5$\,AU; see also \citealt{Weinberg1987}). However, a significant caveat here is that the cloud model used in deriving the estimate \eqref{eq:mc} is extremely crude (a collection of Plummer spheres). A proper understanding of the effect of clouds on wide binaries will only be achieved once we grapple with the complex, filamentary structure of the interstellar medium \citep{sun2022molecular}. In lieu of such an understanding, we take equation \eqref{eq:mc} at face value, and simply ignore the effects of molecular clouds throughout the rest of this work.

The typical timescale for an order-unity change in eccentricity by Galactic tidal torques, $t_\mathrm{Gal}$, is given by (see Figures 5 and 6 of Paper I):
\begin{equation}
    \label{eq:t_sec_numerical}
    t_\mathrm{Gal} \approx 4\,\mathrm{Gyr}\, \left(\frac{\rho_0}{0.1 M_\odot/\mathrm{pc}^{3}} \right)^{-1} \left( \frac{m_b}{M_\odot} \right)^{1/2} \left( \frac{a}{10^4\mathrm{AU}} \right)^{-3/2},
\end{equation}
where $\rho_0$ is the local dynamical density \citep{Widmark2019-wp}. We may compare this with the timescale for a comparable change in eccentricity via diffusive stellar encounters, which, as we will confirm in the next section, is comparable to $\tion$ (equation \eqref{eq:ionize}).
Assuming $\rho_0 \sim nm_\mathrm{p}$, the ratio is 
\begin{align}
    \label{eq:encounters_Tides_Ratio}
    \frac{\tion}{t_\mathrm{Gal}} \approx \frac{0.1}{\ln \Lambda} \frac{m_\mathrm{b}}{m_\mathrm{p}}\frac{\sigma}{v_\mathrm{b}} \approx 1.3\,\left(\frac{\ln \Lambda}{10} \right)^{-1}
    \left(\frac{m_\mathrm{b}}{m_\mathrm{p}}\right)\left( \frac{\sigma}{40 \,\mathrm{km/s}}\right) \left(\frac{a}{10^4\,\mathrm{AU}}\right)^{1/2},
\end{align}
independent of the local density.

\begin{figure}
    \centering
    \includegraphics[width=0.49\textwidth]{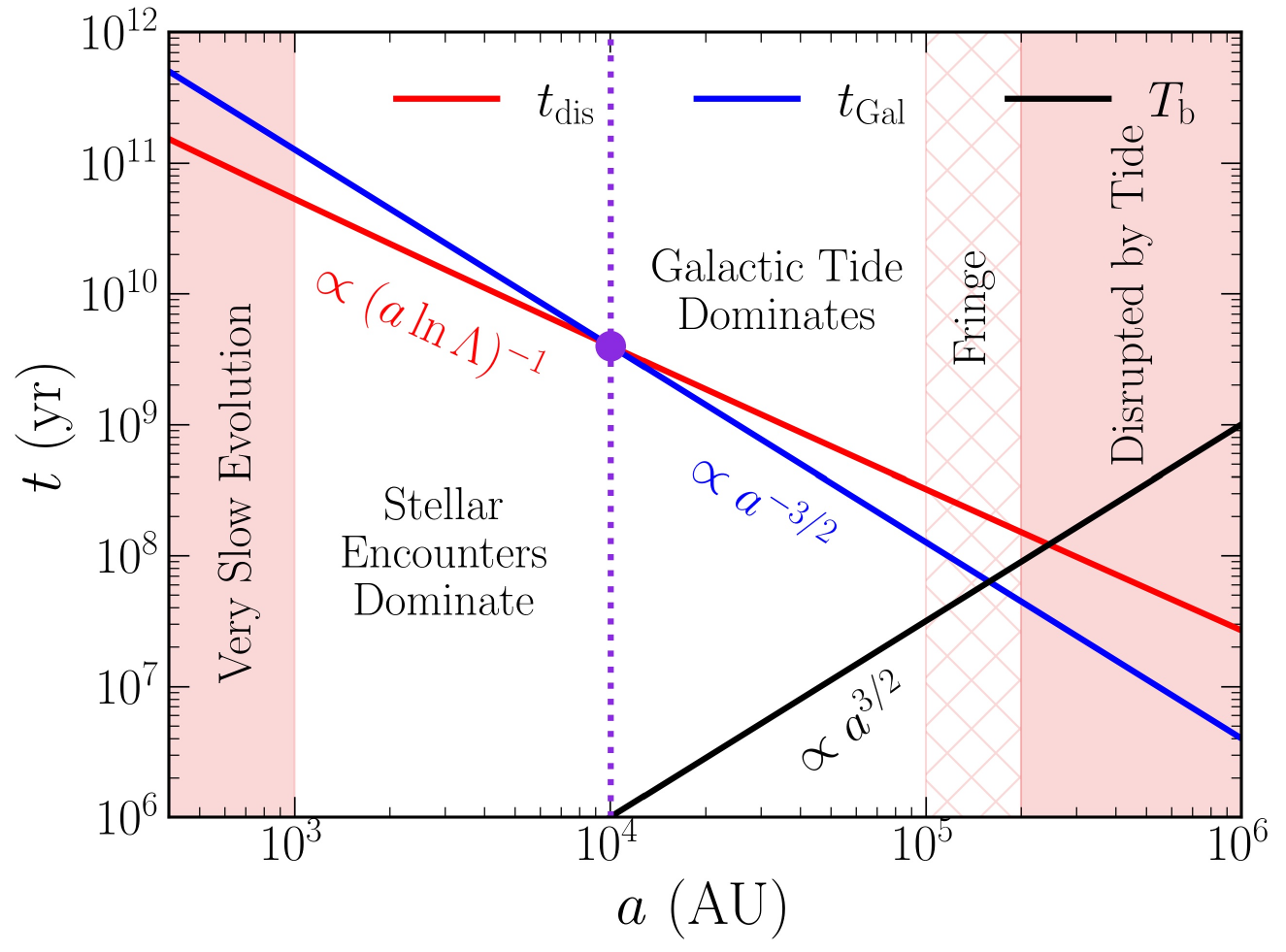}
    \caption{Key timescales determining the dynamical evolution of wide binaries as a function of semimajor axis $a$. The timescale for disruption and eccentricity evolution by stellar encounters $\tion$ (equation \eqref{eq:ionize}) and the timescale for eccentricity evolution driven by Galactic tides $t_\mathrm{Gal}$ (equation \eqref{eq:t_sec_numerical}) are substantially longer than the inner orbital timescale $T_\mathrm{b}$ (equation \eqref{eq:binary_orbit_time}) for all but the widest of binaries (which are quickly disrupted anyway). Galactic tidal torques are the dominant effect for binaries with $a \gtrsim 10^4\,$AU, whereas encounters dominate for $a<10^4$AU. Throughout, we have assumed $n = 0.1/\mathrm{pc}^{3}$, $m_\mathrm{b} = m_\mathrm{p} = 1\Msun$, $\rho_0 = m_\mathrm{p} n$, $\sigma = 40$\,km/s, and $b_\mathrm{max} = a$.}
    \label{fig:timescales}
\end{figure}

In Figure \ref{fig:timescales} we plot the timescales $t_\mathrm{dis}$ and $t_\mathrm{Gal}$ as a function of semimajor axis $a$ using parameters characteristic of wide stellar binaries in the Solar neighborhood (see the caption for details). We see that Galactic tidal torques dominate the evolution for those binaries with $a \gtrsim 10^4 \mathrm{AU}$ whereas stellar encounters dominate for $a < 10^4 \mathrm{AU}$, although the weak scaling $t_\mathrm{dis}/t_\mathrm{Gal} \propto (\ln \Lambda)^{-1} a^{1/2}$ (equation \eqref{eq:encounters_Tides_Ratio}) means that the two effects are not separated by a large factor at any realistic $a$. However, it is worth remembering that most binaries observed with, say, $a \sim 10^4$AU today were actually born with significantly smaller $a$ values, and have become wider over time, as we will see (\S\ref{sec:Present_Day}).  Thus, a large fraction of surviving wide binaries spent much of their lives in a regime where Galactic tidal torques were subdominant compared to stellar encounters. In Paper I we isolated the effects of Galactic tides while ignoring stellar encounters. In this paper we do the opposite: we consider only stellar encounters, and ignore Galactic tides (except at the point where they are required to disrupt binaries, {which we crudely model with a `sink' at} $a = 2\times 10^5$\,AU). We leave a synthesis of the two effects to future work.

Finally, we ignore the fact that some disrupted pairs of stars may be `re-bound' into binaries by stellar encounters \citep{jiang2010evolution}. {This rebinding process affects only the very highest-$a$ tail of the wide binary distribution, which we are already omitting from our final results (see the discussion of the fringe in \S\ref{sec:Approximation_Weak}).}


\section{Fokker-Planck equation}
\label{sec:Kinetic_Theory}

In this section we make use of the assumptions listed in \S\ref{sec:approximations}
in order to derive a Fokker-Planck kinetic equation that describes the evolution of an ensemble of wide binaries. We then illustrate some properties of this equation analytically, and in particular we prove that an arbitrary initial DF will always be driven towards the thermal distribution in eccentricity at each $a$.


\subsection{Distribution function} 

Since we are assuming that binaries are born with isotropic orientations, and that subsequent encounters can be drawn from an isotropic relative velocity distribution (\S\ref{sec:Approximation_Isotropic_Impulsive}), we may ignore any angular dependence in the problem and consider a distribution of binaries only in $(a,e)$ space (which is fundamentally different from the calculation when one includes Galactic tides --- see Paper I). In fact, since the thermal eccentricity distribution is uniform in the \textit{square} of eccentricity at fixed $a$, viz. $P(e) \md e = 2e\, \md e = \md (e^2)$, the natural coordinates with which to formulate the kinetic theory are not $(a,e)$ but rather $(a, \esq)$, where 
\begin{align}
    \esq \equiv e^2.
\end{align}

Let the DF in these coordinates be $f(a, \esq, t)$, such that the number of binaries near location $(a,\esq)$ at time $t$ is proportional to $f(a, \esq, t)\,\md a\, \md \esq$. Then a thermal eccentricity distribution is independent of $\esq$, while a subthermal (superthermal) DF has $\partial f/\partial \esq < 0$ ($ > 0$) for all $\esq$:
\begin{equation}
    \frac{\partial f}{\partial \esq} = 
    \begin{cases}
        < 0  \,\,\,\,\,\,\,\,\,\,\,\,\,\,\,\,\, \mathrm{(subthermal)},
        \\
        \,\,\,\,\,\, 0  \,\,\,\,\,\,\,\,\,\,\,\,\,\,\,\,\, \mathrm{(thermal)},
        \\
        > 0   \,\,\,\,\,\,\,\,\,\,\,\,\,\,\,\,\, \mathrm{(superthermal)}.
    \end{cases}
\end{equation}
As an example, a power-law eccentricity DF $P^{(\alpha)}(e)=(1+\alpha)e^\alpha$ at every $a$ corresponds to 
\begin{equation}
    \label{eq:DF_separable}
    f^{(\alpha)}(a,\esq) = N(a) \tilde{P}^{(\alpha)}(\esq),
\end{equation}
where 
\begin{equation}
    \label{eq:powerlaw_alpha_esq}
    \tilde{P}^{(\alpha)}(\esq)\equiv \frac{\alpha+1}{2} \esq^{(\alpha-1)/2}, \,\,\,\,\,\,\,\, \int_0^1 \md \esq \, \tilde{P}^{(\alpha)}(\esq) = 1,
\end{equation}
and $N(a)$ is an arbitrary semimajor axis distribution.

\begin{figure}
    \centering
    \includegraphics[width=0.49\textwidth]{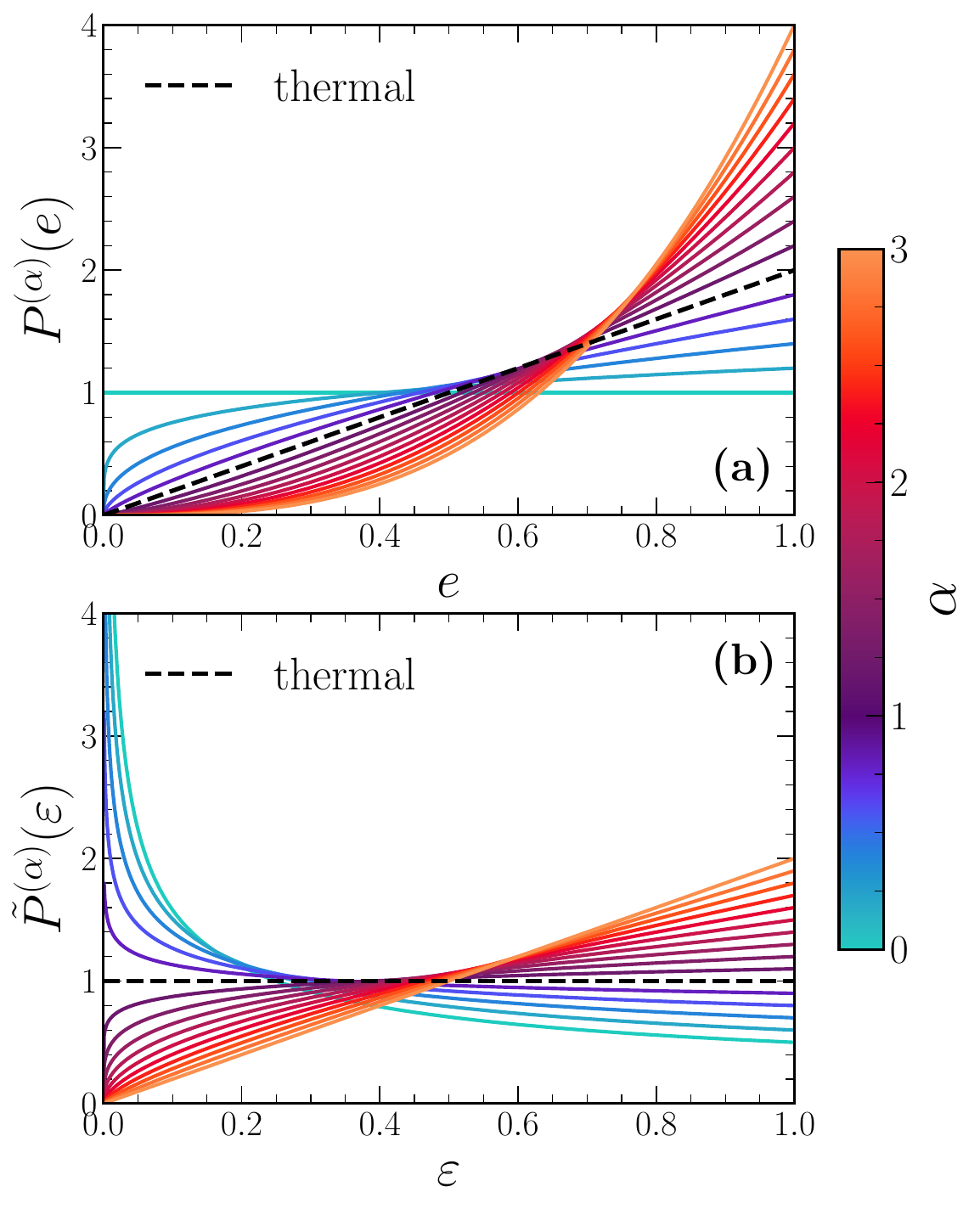}
    \caption{Examples of power-law eccentricity DFs for sixteen evenly spaced values of the power law index $\alpha$ from 0 to 3 inclusive (the same distributions shown in Figure 1 of \citealt{MH23}). Panel (a) shows the eccentricity distribution $P(e) = (1+\alpha)e^\alpha$, 
    while panel (b) shows the corresponding DF in the space of $\esq \equiv e^2$, namely
    $\tilde{P}^{(\alpha)}(\esq) = (\alpha+1)/2 \esq^{(\alpha-1)/2}$. In both panels the black dashed line indicates the thermal distribution ($\alpha = 1$).}
    \label{fig:feps}
\end{figure}

In Figure \ref{fig:feps} we plot both $P(e)$ and $\tilde{P}^{(\alpha)}(\esq)$ for different values of $\alpha$. Subthermal (superthermal) power-law DFs have $\alpha<1$ ($\alpha>1$), while the thermal DF corresponds to $\alpha=1$. Of course, in general, the distribution function need not have the separable form of equation \eqref{eq:DF_separable}.


\subsection{Fokker-Planck equation in ($a, \esq$) space}
\label{sec:FP_in_a_esq}


The fact that stellar encounters are \textit{weak} allows us to treat the evolution of $f$ by means of a Fokker-Planck equation {(but see \S\ref{sec:Approximation_Weak} for caveats)}. For simplicity, we will begin by assuming all binaries are born at $t=0$ (we will discuss the effect of including a more realistic birth history in \S\ref{sec:Source_Sink}). Thus, we are considering the evolution of an initial condition $f(a,\esq, 0)$ under many weak, isotropic encounters. The Fokker-Planck equation in $(a, \esq)$ space that governs this evolution is \citep{cohn1979numerical}:
\begin{align}
    \label{eq:Fokker_a_esq}
    \pder{f}{t} = &
    -\pder{}{a} [\overline{\langle{\Delta a}\rangle} f]
    -\pder{}{\esq} [\overline{\langle{\Delta \esq}\rangle} f] \nn \\
    & 
    + \frac{1}{2}\pder{^2}{a^2} [ \orbitmeansquareDelta{a} f]
    + \frac{1}{2}\pder{^2}{\esq^2} [ \orbitmeansquareDelta{\esq} f]
    \nn \\
    &
    + \frac{\partial^2 }{\partial a \partial \esq} [\overline{\langle{\Delta a \Delta \esq}\rangle} f].
\end{align}
Here, $\meanDelta{a}$, for instance, refers to the total change in $a$ of a binary due to many weak encounters \textit{at a fixed orbital phase} (i.e. fixed $\bm{r}$ and $\bm{v}$),
divided by the time over which those encounters were sampled. The additional overbars, as in $\overline{\meanDelta{a}}$, represent a subsequent average over the Keplerian orbit of the binary at fixed $(a, \esq)$.

In Appendix \ref{sec:Encounter_Average}, we calculate the encounter-averaged quantities $\langle \Delta a \rangle$, $\langle \Delta \esq \rangle$ etc. by assuming the encounters are isotropic, impulsive (\S\ref{sec:Approximation_Isotropic_Impulsive}) and weak (\S\ref{sec:Approximation_Weak}). These encounter-averaged quantities all involve the common factor $\meansquareDelta{\bm{v}}$, as well as various factors of $r^2$ and $v^2$. In general, $\meansquareDelta{\bm{v}}$ itself depends on orbital phase --- for instance, for a highly eccentric binary, distant encounters would be more effective while the binary is at apocenter than when it is at pericenter \citep{penarrubia2019stochastic}. However, under the approximation that all encounters are \textit{penetrative} (\S\ref{sec:Approximation_Penetrative}), each encounter just impinges upon one member of the binary while ignoring the other, and so the orbital phase information is irrelevant. Under this penetrative approximation, we can take $\meansquareDelta{\bm{v}}$ from \cite{jiang2010evolution}'s equation (33), making the simplification that there is only one mass species under consideration, and doubling their answer since kicks to either star are equally effective:
\begin{align}
    \label{eq:Jiang_Delta_V_Squared}
    \meansquareDelta{\bm{v}} = \frac{16\sqrt{2\pi}G^2 m_\mathrm{p}^2 n \ln \Lambda}{\sigma}.
\end{align}
Here, 
\begin{equation}
    \ln\Lambda \equiv \ln\left(\frac{b_\mathrm{max} \sigma^2}{G(m_\mathrm{b}/2 + m_\mathrm{p})}\right) \sim \ln\left(\frac{\sigma^2}{v_\mathrm{b}^2}\right),
    \label{eq:Coulomb}
\end{equation}
is the Coulomb logarithm, and we set $b_\mathrm{max} = a$ as the maximum impact parameter at which the penetrative approximation is valid; then $v_\mathrm{b} \sim (Gm_\mathrm{b}/a)^{1/2}$ is the binary's typical orbital speed (and we have assumed $m_\mathrm{p} \sim m_\mathrm{b}$). Note also that by comparing with equation \eqref{eq:ionize}, we have $\meansquareDelta{\bm{v}} \sim v_\mathrm{b}^2/\tion(a)$.

The right hand side of \eqref{eq:Jiang_Delta_V_Squared} depends on the binary's orbital elements only through the semimajor axis $a$, and even this dependence is very weak (only logarithmic). Crucially, \eqref{eq:Jiang_Delta_V_Squared} does \textit{not} depend on orbital phase. This lack of phase-dependence allows us to orbit-average the coefficients $\langle \Delta a \rangle \to \overline{\langle \Delta a \rangle}$, $\, \langle \Delta \esq  \rangle\to\overline{\langle \Delta \esq  \rangle}$, etc. by substituting the expressions for $\overline{r^2}$ and $\overline{v^2}$ given in Appendix \ref{sec:orbavg}. Putting the results of Appendices \ref{sec:Encounter_Average} and \ref{sec:orbavg} together, we find that the coefficients entering the Fokker-Planck equation \eqref{eq:Fokker_a_esq} are as follows.

First, the drift and diffusion coefficients in semimajor axis are given by
\begin{align}
    \label{eq:a_Drift_orbitavg}
    & \orbitmeanDelta{a} = \frac{7\meansquareDelta{\bm{v}}}{3Gm_\mathrm{b}} a^2 \equiv A_a(a), \\
    \label{eq:a_Diffusion_orbitavg}
    & \orbitmeansquareDelta{a} = \frac{4\meansquareDelta{\bm{v}}}{3Gm_\mathrm{b}} a^3 \equiv D_a(a).
\end{align}
The drift coefficient \eqref{eq:a_Drift_orbitavg} is positive definite, meaning soft binaries get softer;
the naive timescale one derives from this coefficient, $t_a \sim a/\orbitmeanDelta{a}$, 
is essentially the same as the disruption timescale formula \eqref{eq:ionize_numerical}.
The diffusion coefficient \eqref{eq:a_Diffusion_orbitavg} is also positive definite, as it must be. Note that neither of these coefficients depend on $\esq$, whereas outside of the penetrative regime they generally would depend on $\esq$ \citep{penarrubia2019stochastic}.

Second, the drift and diffusion coefficients in $\esq$ are
\begin{align}
   \label{eq:epsilon_drift_avg}
   & \orbitmeanDelta{\esq} = \frac{5\meansquareDelta{\bm{v}}}{3Gm_\mathrm{b}} a(1-2\esq) \equiv A_\esq(a, \esq), \\
   \label{eq:epsilon_diff_avg}
   & \orbitmeansquareDelta{\esq} = \frac{10 \meansquareDelta{\bm{v}}}{3Gm_\mathrm{b}} a\, \esq (1-\esq) \equiv D_\esq(a, \esq).
\end{align}
Note that the drift coefficient \eqref{eq:epsilon_drift_avg} is positive for $\esq<1/2$ (meaning near-circular binaries on average become more radial) and negative for $\esq > 1/2$ (meaning near-radial binaries become more circular). It is zero for $\esq = 1/2$, i.e. for $e = 1/\sqrt{2}$. The naive timescale for changes in eccentricity we infer from this coefficient, $t_e \sim \esq/\vert \orbitmeanDelta{\esq} \vert$, is (for most $\esq$ values) of the same order as equation \eqref{eq:ionize_numerical}, i.e. 
\begin{align}
t_e \sim t_a\sim t_\mathrm{dis},
\label{eqn:timescales_equivalent}
\end{align}
which justifies the claim we made in deriving equation \eqref{eq:encounters_Tides_Ratio}.\footnote{{Interestingly, \cite{penarrubia2019stochastic} found the same balance of timescales \eqref{eqn:timescales_equivalent}, even though he treated the case where distant encounters dominate and penetrative encounters are negligible. The relation \eqref{eqn:timescales_equivalent} seems to be a universal property of (not necessarily Keplerian) stellar systems evolving under impulsive perturbations.}} Meanwhile, the diffusion coefficient \eqref{eq:epsilon_diff_avg} is positive except at $\esq=0$ and $\esq=1$, where it is zero, guaranteeing no unphysical flux of binaries across the $e=0$ and $e=1$ boundaries. The $\esq$ dependence of the coefficients \eqref{eq:epsilon_drift_avg}-\eqref{eq:epsilon_diff_avg} already suggests that any DF will be driven towards uniformity in $\esq$, i.e. thermal in eccentricity; we prove this formally in \S\ref{sec:Thermal_Properties} and Appendix \ref{sec:Casimir_proof}. For later use we also note the following identity (see also \citealt{magorrian1999rates}):
\begin{equation}
    \label{eq:not_fluctdiss}
   A_\esq  = \frac{1}{2}\frac{\partial D_\varepsilon}{\partial\varepsilon}.
\end{equation}

Third, kicks in $a$ and $\esq$ are uncorrelated:
\begin{align}
   \label{eq:a_esq_correlator}
    \overline{\langle \Delta a \Delta \esq \rangle} = 0.
\end{align}
This identity allows for a very simple numerical solution to the Fokker-Planck equation (\S\ref{sec:Simulation_Method}), which is another advantage of using $(a,\esq)$ as phase space coordinates rather than $(E, J)$ or $(a, e)$.

The identities \eqref{eq:not_fluctdiss} and \eqref{eq:a_esq_correlator} allow us to simplify the Fokker-Planck equation \eqref{eq:Fokker_a_esq} so that it reads
\begin{align}
    \label{eq:Fokker_a_esq_simpler}
    \frac{\partial  f}{\partial t} = - \frac{\partial }{\partial a}\left[ A_a  f - \frac{1}{2}\frac{\partial }{\partial a} (D_a  f)\right] +  \frac{1}{2} \frac{\partial }{\partial \varepsilon} \left[ D_\varepsilon \frac{\partial  f}{\partial \varepsilon}\right].
\end{align}
{As discussed above, we ignore the complexities of the fringe (\S\ref{sec:Approximation_Weak}) and any re-binding processes (\S\ref{sec:molecular_clouds}), and simply
apply to equation \eqref{eq:Fokker_a_esq_simpler} the crude boundary condition}
\begin{equation}
    f(a_\mathrm{dis}, \esq, t) = 0, \,\,\,\,\,\,\,\,\, \forall \,\esq,\, t,
    \label{eq:disrupting_boundary}
\end{equation}
meaning that binaries reaching semimajor axes of $a_\mathrm{dis} \equiv 2\times 10^5$\,AU get disrupted and hence removed from consideration (e.g. \citealt{jiang2010evolution}).
For simplicity we use the same $a_\mathrm{dis}$ value at all $\esq$ even though in reality the disruption process is (weakly) eccentricity-dependent.


\subsubsection{Power-law DFs}
\label{sec:power_law_DFs}


To gain some intuition about the evolution of binary ensembles under the Fokker-Planck equation \eqref{eq:Fokker_a_esq_simpler}, we rewrite it as
\begin{equation}
    \frac{\partial f}{\partial t } = -\left( \frac{\partial}{\partial a}, \frac{\partial}{\partial \esq} \right) \cdot (f\mathbf{U}),
\end{equation}
where $\mathbf{U} = (U_a, U_\esq)$ is the `velocity' of the phase space flow:
\begin{align}
    U_a = A_a - \frac{1}{2f} \frac{\partial}{\partial a}(D_a f), \\
    U_\esq = A_\esq  - \frac{1}{2f} \frac{\partial}{\partial \esq}(D_\esq f).
\end{align}
If we specialize to the case of independent power laws of index $\alpha$ in eccentricity and $\beta$ in semimajor axis, $f\propto \varepsilon^{(\alpha-1)/2}a^\beta$ (see equation \eqref{eq:DF_separable}), we can explicitly calculate
\begin{align}
    \label{eq:U_a}
    U_a & = \frac{32\sqrt{2\pi}}{3}\frac{Gm_\mathrm{p}^2n}{m_b\sigma}a^2\left[\frac{1-2\beta}{2}\ln\Lambda - 1\right], \\
    \label{eq:U_a_tion}
    & \approx 0.5 \left[\frac{1-2\beta}{2} - \frac{1}{\ln\Lambda}\right] \frac{a}{\tion(a)}, \\
    \label{eq:U_esq}
    U_\esq & = \frac{40\sqrt{2\pi}}{3}\frac{Gm_\mathrm{p}^2n}{m_b\sigma}(1-\alpha)(1-\esq)a\ln\Lambda, \\
    \label{eq:U_esq_tion}
    & \approx 0.7 (1-\alpha)\frac{1-\esq}{\tion(a)},
\end{align}
where in equations \eqref{eq:U_a_tion} and \eqref{eq:U_esq_tion} we have used equation \eqref{eq:ionize} to highlight the dependence of the flow velocity on the disruption timescale.

\begin{figure*}
    \centering
    \includegraphics[width=0.88\textwidth]{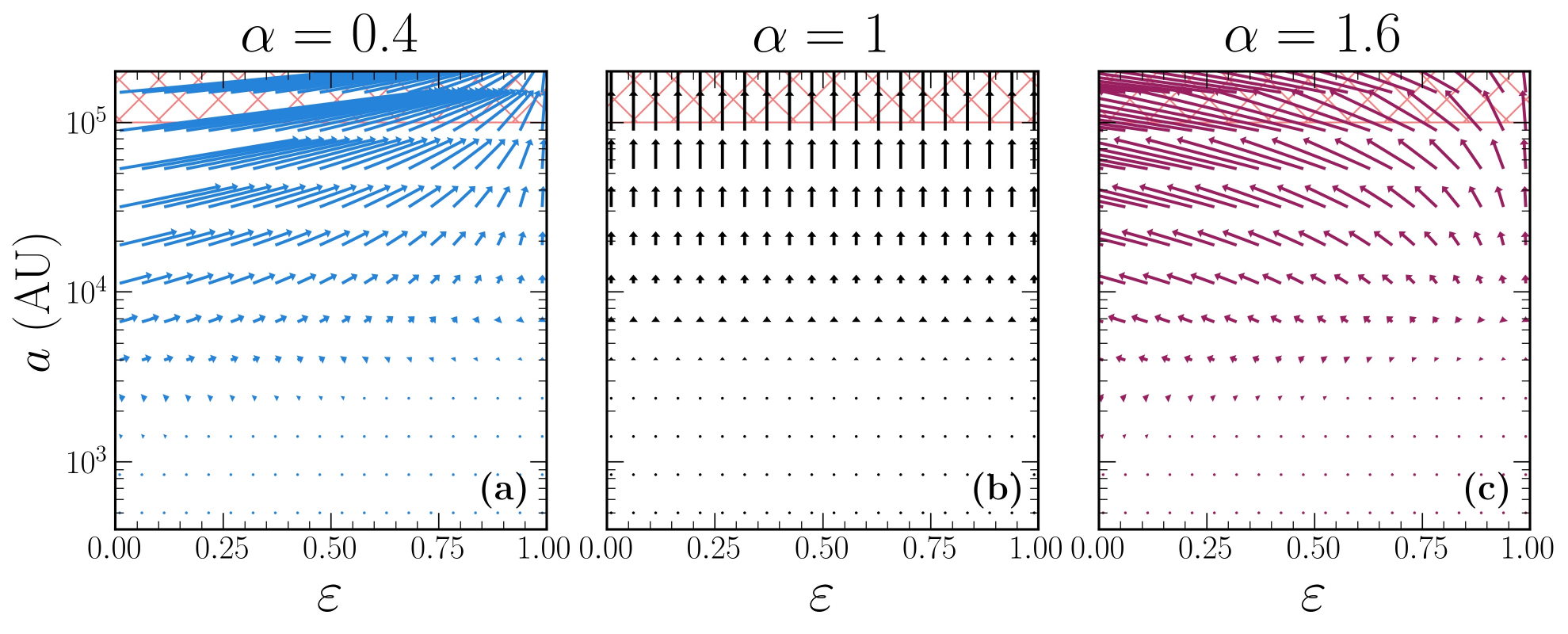}
    \caption{The ``velocity'' of the phase space flow $\mathbf{U} = (U_\esq, U_a)$ given by equations \eqref{eq:U_a} and \eqref{eq:U_esq} for distribution functions following Opik's law ($\beta = -1$) in semimajor axis from $a_\mathrm{min} = 400\,$AU to $a_\mathrm{max} = 2\times 10^5\,$AU and a power law of index $\alpha$ in eccentricity. The length of each arrow is proportional to the amplitude of $\vert \mathbf{U}\vert $. {As in Figure \ref{fig:timescales}, the `fringe' region $a > 10^5\,$AU is shown with red hatches.}}
    \label{fig:velocity_vectors}
\end{figure*}

In Figure \ref{fig:velocity_vectors} we plot the vector field $\mathbf{U}$ in the $(\varepsilon, a)$ plane for fixed $\beta = -1$ (so the binaries follow Opik's law in semimajor axis) and several values of $\alpha$. We see that for the thermal eccentricity distribution (panel (b), with $\alpha=1$) the flux is purely vertical, i.e. towards larger semimajor axes with no change in $\esq$, and that the flow is faster at larger $a$ (as expected from equation \eqref{eq:ionize}). Meanwhile in the other panels, for subthermal (superthermal) DFs, i.e. those with $\alpha < 1$ ($> 1$), the flow has a vertical component but also points to the right (left), so that at fixed $a$ the eccentricity DF tends closer to thermal. Though the velocity field in panel (c) has a leftward-pointing component at $\esq=0$, there is of course no unphysical flux of binaries across this boundary, since the product $f U_\esq \propto \esq^{0.3}(1-\esq)$ vanishes there. 


\subsection{Analytic properties of the Fokker-Planck equation}
\label{sec:Kinetic_Theory_properties}


In general, it is not possible to solve the Fokker-Planck equation \eqref{eq:Fokker_a_esq_simpler} analytically.  Nevertheless, there are several insights we can gain analytically  before turning to a full numerical solution (\S\ref{sec:Numerical_Solution}).

Let us begin by defining the thermal ($\esq$-independent) part of the DF, $F(a,t)$, as
\begin{equation}
    F(a, t) \equiv \int_0^1 \md\varepsilon \, f(a, \varepsilon, t).
\end{equation}
Then without loss of generality we can always write $f$ as a thermal part plus a non-thermal part:
\begin{equation}
    f(a, \esq, t) = F(a, t ) + \delta f(a, \varepsilon, t),
\end{equation}
where
\begin{equation}
    \int \md \esq \, \delta f(a,\esq,t) = 0, \,\,\,\,\,\,\,\,\, \forall \, a, \, t.
\end{equation}
Note that aside from this requirement, $\delta f$ need not be in any sense small compared to $F$.

Obviously, $F$ is just the semimajor axis distribution. Indeed, integrating the Fokker-Planck equation \eqref{eq:Fokker_a_esq_simpler} over $\esq$ tells us how $F$ evolves in time:
\begin{equation}
    \label{eq:mean_evolution}
    \frac{\partial F}{\partial t} = - \frac{\partial }{\partial a}\left[ A_a F - \frac{1}{2}\frac{\partial }{\partial a} (D_a F)\right],
\end{equation}
which is completely independent of $\esq$. In other words, in order to follow the evolution of the thermal component (i.e. the semimajor axis distribution alone), we can ignore eccentricities altogether. In fact, if one ignores the $a$-dependence of the Coulomb logarithm $\ln \Lambda$, then equation \eqref{eq:mean_evolution} actually admits an explicit analytic solution \citep{Weinberg1987,jiang2010evolution}. However, this analytic formula is not very enlightening, and since in this paper we are mostly interested in eccentricity distributions anyway, we will refrain from using it.

We can also get an equation for the evolution of the non-thermal component $\delta f$
by subtracting \eqref{eq:mean_evolution} from \eqref{eq:Fokker_a_esq_simpler}:
\begin{align}
    \label{eq:fluctuation_evolution}
       \frac{\partial \delta f}{\partial t} = - \frac{\partial }{\partial a}\left[ A_a \delta f - \frac{1}{2}\frac{\partial }{\partial a} (D_a \delta f)\right] +  \frac{1}{2} \frac{\partial }{\partial \varepsilon} \left[ D_\varepsilon \frac{\partial \delta f}{\partial \varepsilon}\right].
\end{align}


\subsubsection{Properties of thermal eccentricity DFs}
\label{sec:Thermal_Properties}


Equations \eqref{eq:mean_evolution} and \eqref{eq:fluctuation_evolution} allow us to prove two important results about wide binaries under weak, isotropic, impulsive, penetrative stellar encounters:
\begin{itemize}
    \item \textit{The thermal eccentricity DF is stationary}, in the sense that an initially thermal eccentricity DF remains thermal forever. This is easy to prove: we have $\delta f(a, \esq, t = 0) = 0$, so the right hand side of \eqref{eq:fluctuation_evolution} vanishes, and a non-zero $\delta f$ can never develop.  $\square$
    \item \textit{The thermal eccentricity DF is a unique attractor}, in the sense that an arbitrary eccentricity distribution always tends towards a thermal distribution, despite the fact that the binaries are always drifting towards larger semimajor axis (recall that a thermal distribution of eccentricities is uniform in phase space at \textit{fixed} semimajor axis). To prove this, we investigate the quantity
    \begin{equation}
        C(t) \equiv \int \md a \, \md \esq \, \vert \delta f(a, \esq, t) \vert^2,
        \label{eq:Casimir}
    \end{equation}
    which is analogous to the Casimir $\delta C_2$ that we introduced in Paper I. In Appendix \ref{sec:Casimir_proof} we prove that $C$ is monotonically decreasing in time:
    \begin{equation}
        \label{eq:dCdt_negative}
        \frac{\md C}{\md t} \leq 0,
    \end{equation}
    with equality only once the thermal distribution is achieved ($\delta f = 0$ at all $(a, \esq)$). It follows that in the absence of sources (see \S\ref{sec:Source_Sink}), all solutions to the Fokker-Planck equation \eqref{eq:Fokker_a_esq} tend towards thermal in eccentricity at \textit{every} $a$. $\square$\,\footnote{Note that unlike in Paper I, here we are not able to prove rigorously that the eccentricity distribution at fixed $a$ can never transition, \textit{at any time}, from subthermal to superthermal or vice versa. We are only able to show that  it will inevitably tend towards thermal in the time-asymptotic limit. However, the numerical examples shown in \S\ref{sec:Numerical_Solution} confirm that in reality the `thermal barrier' is indeed never crossed.}
\end{itemize}


\subsubsection{Including binaries born after $t=0$}
\label{sec:Source_Sink}


Up to now we have assumed that all of our binaries were `born' at $t=0$. However, in reality wide binaries do not all form at once but instead have some (largely unknown) birth history. To account for this birth history, one may formally add to the right hand side of \eqref{eq:Fokker_a_esq_simpler} a source term $S(a, \esq, t)$, such that $S \,\md t \,\md a \,\md \esq$ is proportional to the number of binaries being born in the time interval $(t, t+\md t)$ with semimajor axes $\in (a, a+\md a)$ and squared eccentricities $\in (\esq, \esq+\md \esq)$. However, in practice this is not really necessary, because it simply amounts to replacing an equation of the form $\mathcal{L}f(t)=0$, where $\mathcal{L}$ is a time-independent linear operator, with an equation of the form $\mathcal{L}f(t) = S(t)$. One can solve this equation by calculating the Green's function of $\mathcal{L}$ --- which is equivalent to solving the initial value problem with all binaries born at $t=0$ --- and then writing $f$ as the convolution of $S$ with this Green's function \citep{Weinberg1987}.

Physically, such a solution works because each binary in our ensemble is considered to evolve independently of all the others, so for an arbitrary birth history we can always follow the evolution of each `generation' of binaries individually, and then superpose the results at whatever time $t$ we are interested in. The conclusions derived earlier in this section based upon the properties of equation \eqref{eq:Fokker_a_esq_simpler} will then hold for each coeval ensemble \textit{individually}, but do not necessarily hold for the entire distribution taken together. For instance, once a particular generation of binaries has been born, its eccentricity DF will tend towards thermal at all $a$ (\S\ref{sec:Thermal_Properties}) but if there has recently been an extremely rapid burst of high-eccentricity binary formation, then this may have temporarily tipped the overall binary distribution \textit{away} from thermal. The evolution of the eccentricity DF is therefore in principle highly dependent upon what we choose for $S(t)$. In \S\ref{sec:Present_Day} we will investigate an astrophysically relevant example in which binaries are born at roughly a constant rate over $10$ Gyr.

\subsection{Numerical examples}
\label{sec:Numerical_Solution}


We now turn to a fully numerical solution to the Fokker-Planck equation in $(a,\esq)$ space for several example initial distributions.


\subsubsection{Monte-Carlo solution method}
\label{sec:Simulation_Method}


As is well known, one way to solve the Fokker-Planck equation is by Monte-Carlo sampling a large number of Langevin particles, each of which moves during a small timestep $\delta t$ according to the Euler-Maruyama update rules $a \mapsto a + \delta a$, $\varepsilon \mapsto \varepsilon + \delta \varepsilon$ where
\begin{align}
    \delta a & = A_a\delta t + \xi_a\left(D_a \delta t\right)^{1/2} \\
    \delta \varepsilon & = A_{\esq}\delta t + \xi_\varepsilon\left(D_{\varepsilon}\delta t\right)^{1/2}
\end{align}
and $\xi_a$ and $\xi_\varepsilon$ are Gaussian random variables with mean 0 and variance 1 (\citealt{spitzer1987, Numerical_SDEs})\footnote{The fact that kicks in $a$ and $\esq$ are uncorrelated (equation \eqref{eq:a_esq_correlator}) means that the Gaussian random variables $\xi_a$ and $\xi_\epsilon$ can be sampled independently. This would not be the case had we used $(a,e)$, for example.}. We use this method throughout the rest of the paper to solve equation \eqref{eq:Fokker_a_esq_simpler}. In particular, we always use a fixed timestep of $\delta t = 2\times 10^{-5}\,$Gyr so that typical values of $\vert\delta a/a\vert$ and $\vert\delta \esq/\esq\vert$ are $\ll 1$ across all semimajor axes and eccentricities considered to ensure appropriate numerical convergence. In computing the drift and diffusion coefficients, we always assume a maximum impact parameter of $b_\mathrm{max} = a$, a total binary mass of $m_b = 1\Msun$, a perturber mass of $m_\mathrm{p} = 1\Msun$, a perturber number density of $n = 0.1 /\mathrm{pc}^{3}$, and a perturber velocity dispersion of $\sigma = 40\,\mathrm{km/s}$. Unless otherwise stated we run all simulations for $10$\,Gyr, and we always employ $10^5$ particles (which we refer to as `binaries' from now on, although strictly speaking they are just realizations of the stochastic process that underlies the Fokker-Planck equation). We remove from the simulation any binary which achieves a semimajor axis greater than $a_\mathrm{dis} = 2\times 10^5\,$AU, consistent with the boundary condition \eqref{eq:disrupting_boundary}.


\subsubsection{Qualitative illustration of behavior in $(a,\esq)$ space}
\label{sec:sim_setup}


\begin{figure*}
    \centering
    \includegraphics[width=0.83\textwidth]{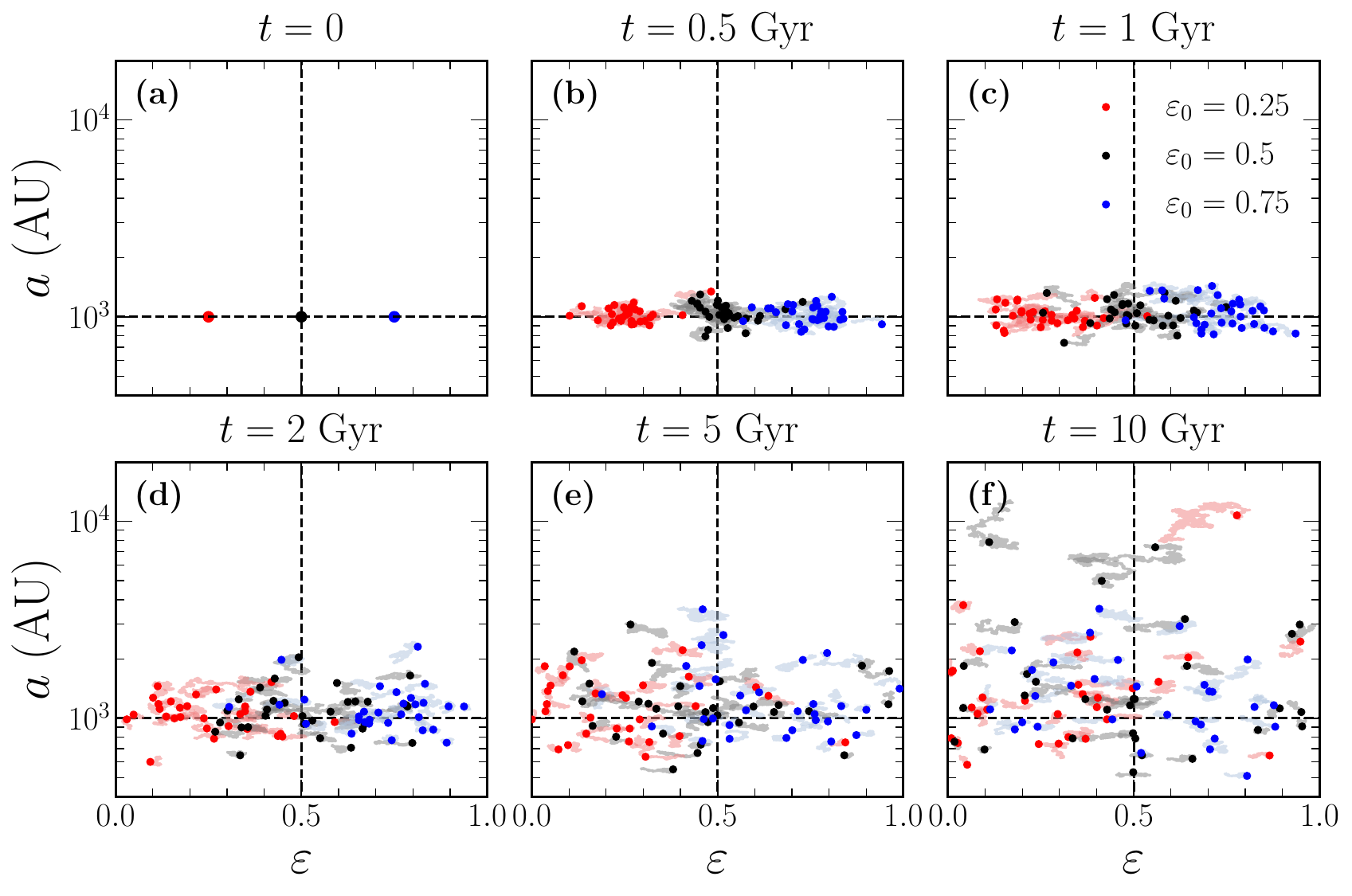}
    \caption{Trajectories for 90 binaries with initial semimajor axes $a_0 = 10^3\,$AU. There are 30 binaries initially located at $\esq_0 = 0.25$ (red), 30 at $\esq_0 = 0.5$ (black) and 30 at $\esq_0 = 0.75$ (blue). The points indicate each binary's position in the $(\esq,a)$ plane at each time, while the lighter-colored tail behind each point shows the trajectory of the binary in the preceding $0.2$\,Gyr.}
    \label{fig:a_epsilon_trajectories}
\end{figure*}
\begin{figure*}
    \centering
    \includegraphics[width=0.8\textwidth]{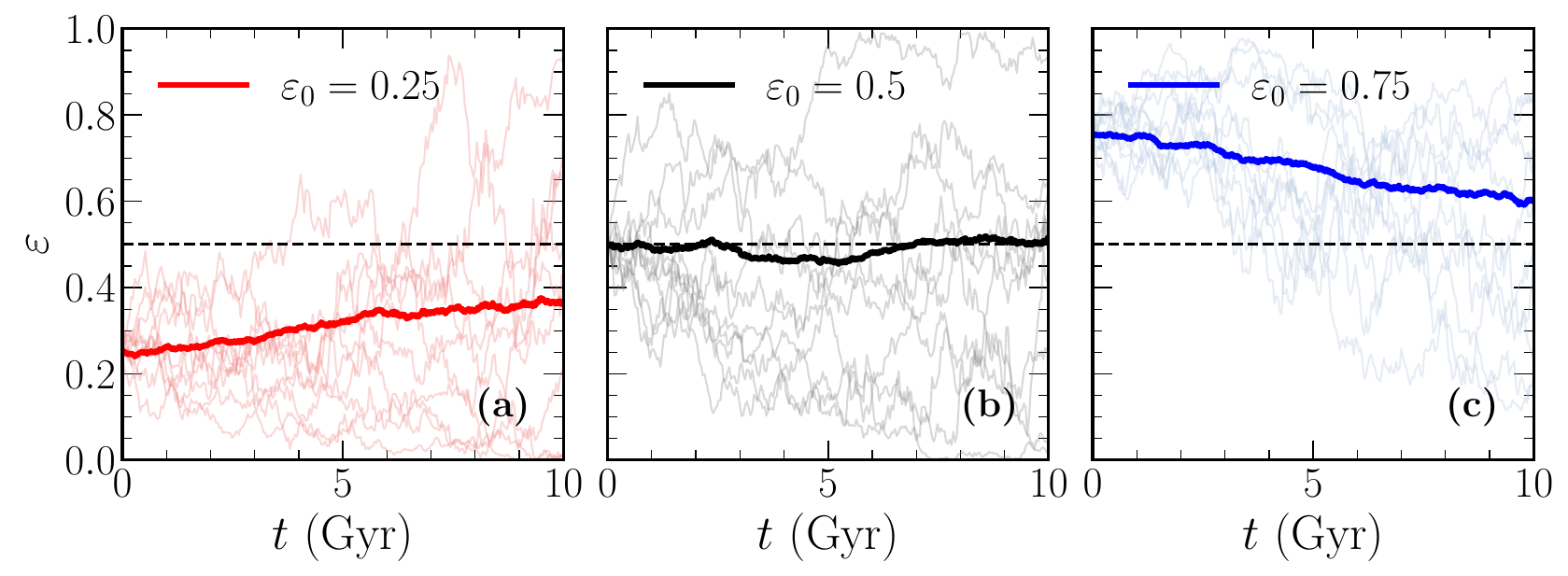}
    \caption{In each panel, light colors show the $\varepsilon(t)$ timeseries for $10$ randomly selected binaries from Figure \ref{fig:a_epsilon_trajectories} (with the same color coding). The thick, darker colored lines show the average $ \esq(t)$ over all $30$ binaries of the given color from Figure \ref{fig:a_epsilon_trajectories}. As anticipated by equation \eqref{eq:epsilon_drift_avg}, binaries with $\varepsilon_0 = 0.5$ (middle panel) do not exhibit a significant drift on average, while those with $\varepsilon_0 \neq 0.5$ (left and right panels) drift toward $\varepsilon = 0.5$.}
    \label{fig:epsilon_trajectories}
\end{figure*}

To illustrate the characteristic evolution that is driven by the Fokker-Planck equation \eqref{eq:Fokker_a_esq_simpler}, we first turn to a numerical solution for some extremely simple initial conditions, namely 
\begin{equation}
    f(a,\esq,0) \propto \delta(a-a_0)\delta(\esq-\esq_0),
\end{equation}
for certain choices of $a_0$ and $\esq_0$. In Figure \ref{fig:a_epsilon_trajectories} we show the resulting evolution of just 90 binaries through the $(\esq, a)$ plane. All binaries are initialized with $a_0 = 10^3$AU; 30 of them have $\esq_0 = 0.25$ (shown with red dots), 30 have $\esq_0 = 0.5$ (black dots) and 30 have $\esq_0 = 0.75$ (blue dots). The different panels (a)-(f) show the subsequent locations of these binaries in $(\esq, a)$ at different times $t$, and the lightly shaded `tail' behind each particle shows its trajectory throughout the preceding $0.2$ Gyr. As expected, the binaries gradually diffuse in both $a$ and $\esq$, and there is a tendency for them to drift towards higher $a$, owing to the positive-definite value of $A_a$ (equation \eqref{eq:a_Drift_orbitavg}). By $t=10$ Gyr (panel (f)) the binaries are reasonably well spread out in $\esq$, although the initial condition $\propto \delta(\esq - \esq_0)$ has clearly not yet been erased (red binaries still tend to be found at lower $\esq$, and blue binaries at higher $\esq$).

To demonstrate the $\esq$-behavior further we present Figure \ref{fig:epsilon_trajectories}, each panel of which shows the $\esq(t)$ time series for $10$ randomly chosen binaries from Figure \ref{fig:a_epsilon_trajectories} (with the same color coding), over the full $10$ Gyr for which we ran the simulation. In each panel of Figure \ref{fig:epsilon_trajectories} we also superimpose with thick lines the average $ \esq(t) $ value over all 30 binaries from the corresponding initial-$\esq_0$ subsample. We notice that while the average of both red and blue ensembles tends over time towards $0.5$ (reflecting the binaries' tendency to be driven towards a thermal distribution, see \S\ref{sec:Thermal_Properties}), it has not yet converged to this value after $10$ Gyr.  This is consistent with the estimate \eqref{eq:ionize_numerical} which says that an ensemble of $a=10^3$AU binaries takes significantly longer than a Hubble time to disrupt (as we will confirm below, the disruption and thermalization timescales are essentially the same). Had we shown a similar example with all binaries born at $a_0 = 10^4$AU, the convergence would have been much faster.

 
\subsubsection{Semimajor axis DF evolution for coeval ensembles}
\label{sec:Numerical_SemimajorAxes}


As shown in \S\ref{sec:Kinetic_Theory_properties}, the evolution of the semimajor axis distribution is independent of eccentricity, so to follow it we need only solve equation \eqref{eq:mean_evolution} subject to the boundary condition \eqref{eq:disrupting_boundary}. We also argued in \S\ref{sec:Source_Sink} that to follow the evolution of a set of binaries with an arbitrary birth history, we need only superpose the results of coeval ensembles. Let us therefore study the evolution of various coeval ensembles, before considering a less trivial birth history in \S\ref{sec:Present_Day}.

We solved equation \eqref{eq:mean_evolution} for 5 different initial conditions, each starting with $10^5$ binaries. In the first four cases we used an initial condition $F(a,0) \propto \delta(a-a_0)$ where $a_0$ took the values  (a) $10^3$\,AU, (b) $10^{3.5}$\,AU, (c) $10^{4}$\,AU and (d) $10^{4.5}$\,AU. In the fifth case we took $F(a,0) \propto a^{-1}$ (Opik's law)
with $a \in (a_\mathrm{min}, a_\mathrm{dis})$, setting $a_\mathrm{min} = 400$\,AU.

\begin{figure}
    \centering
    \includegraphics[width=0.45\textwidth]{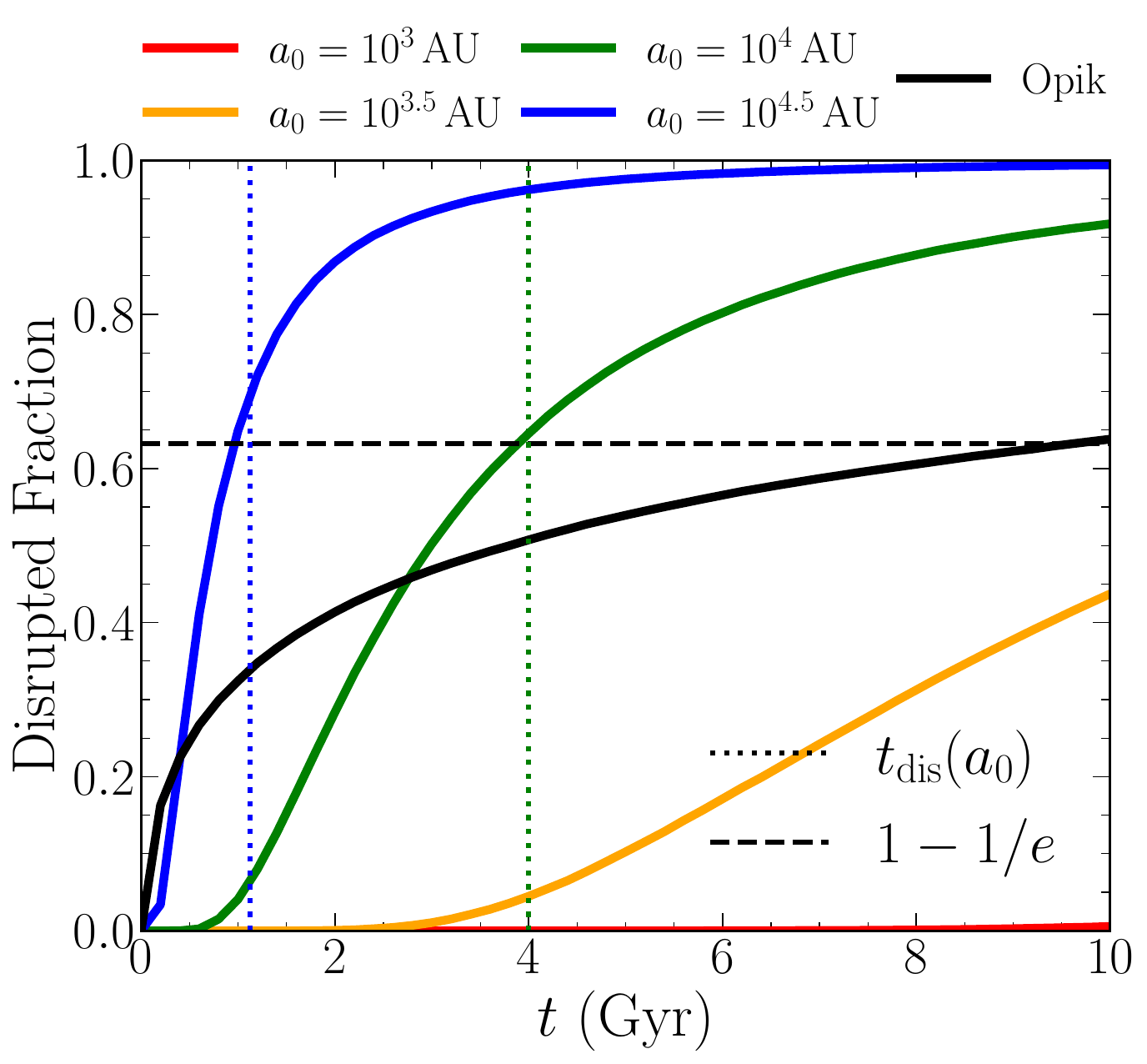}
    \caption{The fraction of binaries that are disrupted over the course of $10$\,Gyr, found by solving equation \eqref{eq:mean_evolution} numerically with a sink at $a=2\times 10^5$AU (see \S\ref{sec:FP_in_a_esq} for caveats). The solid lines correspond to the simulations with varied initial semimajor axis distributions, while the vertical dotted lines denote the predicted disruption timescale $\tion$ from equation \eqref{eq:ionize_numerical}, and the horizontal dashed line denotes an disruption fraction of $1 - 1/e$.}
    \label{fig:ionized_fraction}
\end{figure}

In Figure \ref{fig:ionized_fraction} we show the  resulting fraction of  binaries that get disrupted according to equation \eqref{eq:mean_evolution} after time $t$. The vertical dotted lines show the estimate of the disruption timescale \eqref{eq:ionize_numerical} for the cases $a_0 = 10^4$\,AU and $a_0=10^{4.5}$\,AU. We see that these coincide fairly closely with the time at which the disruption fraction reaches $1-1/e \approx 63\%$; this is an accidental consequence of the fact that the survival probability at time $t$ for a wide binary of initial semimajor axis $a_0$ (e.g. \citealt{Weinberg1987}, equation (B19)) is reasonably well-approximated by $\exp[-t/t_\mathrm{dis}(a_0)]$. We see that the vast majority of binaries born with $a \gtrsim 10^4$\,AU do not survive $10$\,Gyr of evolution. This emphasizes the fact that the majority of very wide binaries we see above $a=10^4$\,AU today spent most of their lives with significantly smaller semimajor axes. By comparison, binaries born at $a \sim 10^3$\,AU  tend to live relatively quiet lives and have a high probability of surviving for a Hubble time.

\begin{figure*}
    \centering
    \includegraphics[width=0.99\textwidth]{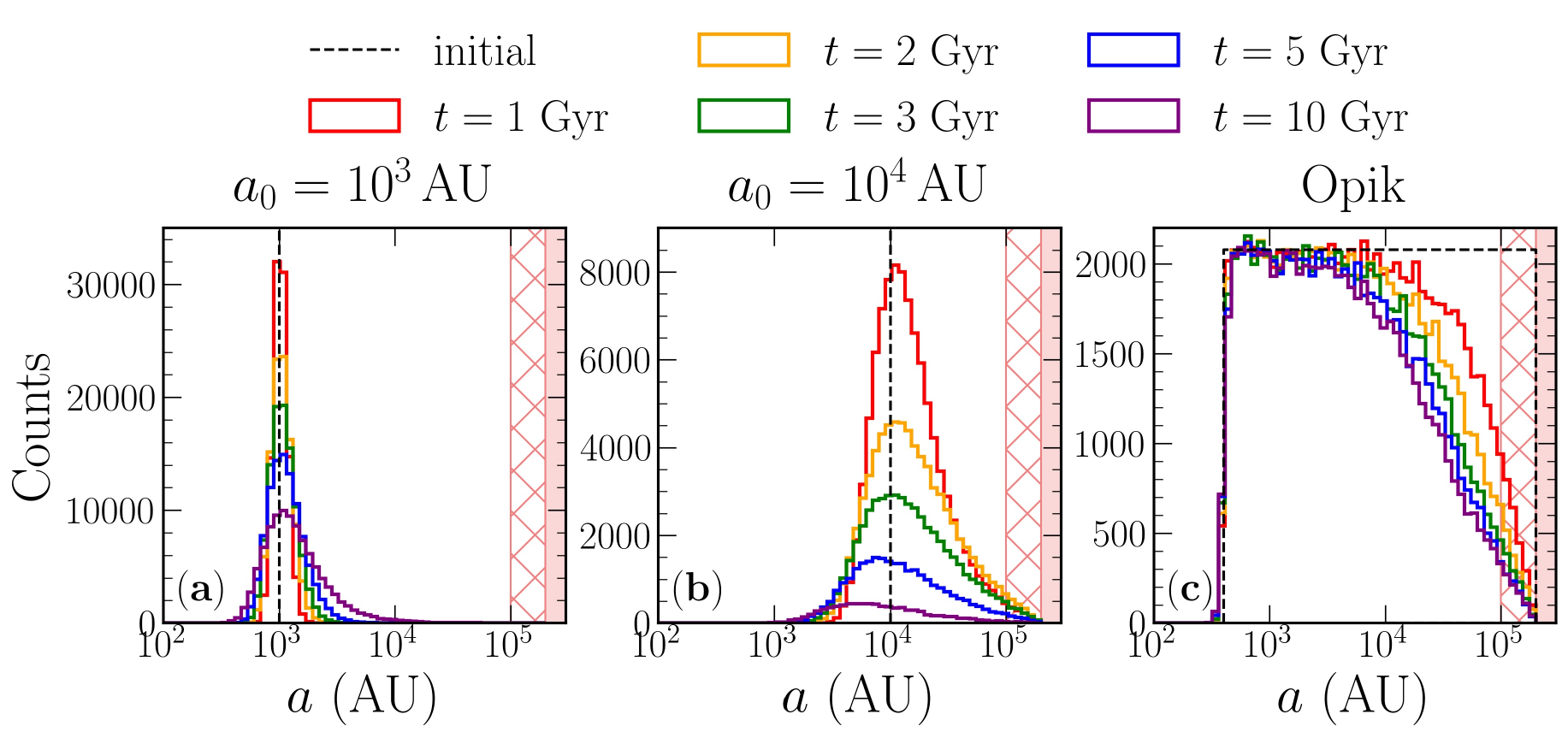}
    \caption{Semimajor axis distributions over time for different choices of the initial distribution, found by solving equation \eqref{eq:mean_evolution} numerically.
    In panels (a) and (b), all $10^5$ binaries are initially located at $a=a_0$ 
    with $a_0 = 10^3\,$AU and $10^4\,$AU respectively. In panel (c) the binaries are initially distributed according to Opik's law between $a_\mathrm{min}=400$ AU and $a_\mathrm{dis}=2\times 10^5$AU. {As in Figure \ref{fig:timescales}, the `fringe' region $a > 10^5\,$AU and tidal disruption region $a > 2\times 10^5\,$AU are indicated with red hatches and solid red shading respectively.}}
    \label{fig:a_histograms}
\end{figure*}

In Figure \ref{fig:a_histograms} we show the resulting histograms in semimajor axis space at various times for three of the initial conditions from Figure \ref{fig:ionized_fraction}, namely (a) $a_0 = 10^3$\,AU, (b) $a_0=10^4$\,AU, and (c) Opik's law. One slightly non-intuitive finding from panel (a) is that the peak of the distribution of surviving binaries actually \textit{decreases} with time. This tendency simply reflects the fact that binaries with larger $a$ evolve faster and thus exit the simulation sooner.


\subsubsection{Eccentricity DF evolution for coeval ensembles}
\label{sec:Eccentricity_DF_Coeval}


We now consider the eccentricity evolution of various ensembles by solving the full Fokker-Planck equation \eqref{eq:Fokker_a_esq_simpler}. For our initial conditions $f(a,\esq,0)$ we use DFs of the separable form \eqref{eq:DF_separable}, which consist of a semimajor axis distribution multiplied by an $\alpha$ power-law eccentricity distribution.

\begin{figure*}
    \centering
    \includegraphics[width=0.95\textwidth]{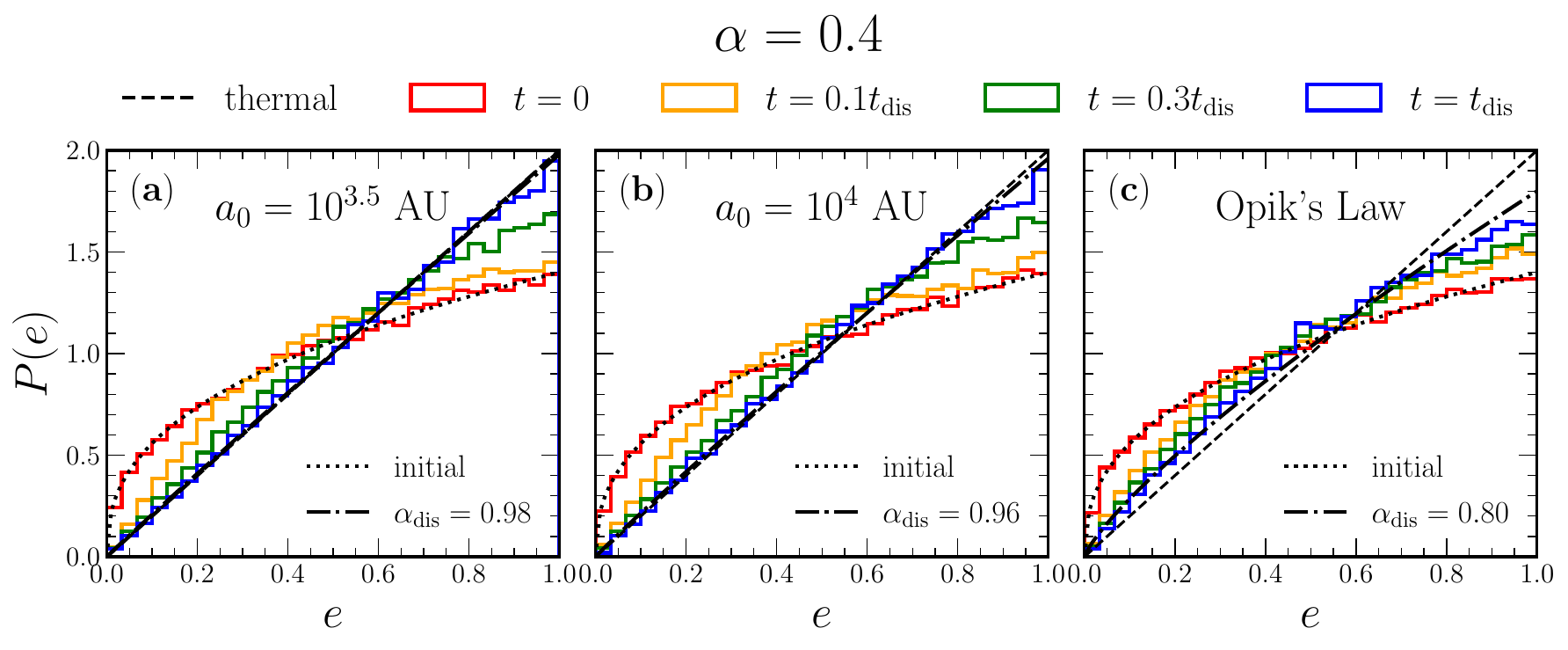}
    \caption{Eccentricity distribution $P(e)$ at different times (different colored lines) for a DF whose initial condition is of the form \eqref{eq:DF_separable} with $\alpha = 0.4$. The initial semimajor axis distribution in panel (a) is a delta-function at $a_0 = 10^{3.5}$\,AU, panel (b) is the same except $a_0 = 10^{4}$\,AU, and panel (c) takes Opik's law with $a_\mathrm{min} = 400$\,AU. In each case time is measured in units of $t_\mathrm{dis}$ (equation \eqref{eq:ionize}); for Opik's law, we define $t_\mathrm{dis}$ to be the time at which the disrupted fraction of binaries reaches $1-1/e = 63\%$ (Figure \ref{fig:ionized_fraction}).}
    \label{fig:eccentricity_DF_alpha_pt4}
\end{figure*}

In Figure \ref{fig:eccentricity_DF_alpha_pt4} we plot $P(e)$ at different times $t$ (different colored lines) for an initial DF with eccentricity power law $\alpha=0.4$ (which is subthermal).  Panel (a) is for a semimajor axis distribution initially given by $\propto \delta(a-a_0)$ with $a_0 = 10^{3.5}$\,AU, panel (b) is the same except $a_0 = 10^{4}$\,AU, and panel (c) uses Opik's law with $a_\mathrm{min} = 400$\,AU. In each panel, the time is measured in units of $t_\mathrm{dis}$, which for panels (a) and (b) is defined using equation \eqref{eq:ionize} with $a=a_0$, and for panel (c) is defined to be the time at which the black (Opik) curve in Figure \ref{fig:ionized_fraction} coincides with $1-1/e$. In each case we only show the simulation results up to $t=t_\mathrm{dis}$, and fit the eccentricity DF at that time with a power law $P(e) \propto e^{\alpha_\mathrm{dis}}$ as indicated in the Figure. We see that in every panel the eccentricity DF is driven towards thermal as expected, although the thermal distribution is never quite achieved by the time $t=t_\mathrm{dis}$. In particular, in panel (c), which uses Opik's law for the initial semimajor axis DF, the eccentricity DF after $t_\mathrm{dis}$ (which is $\sim5$ Gyr, see Figure \ref{fig:ionized_fraction}) is still strongly subthermal.

\begin{figure*}
    \centering
    \includegraphics[width=0.95\textwidth]{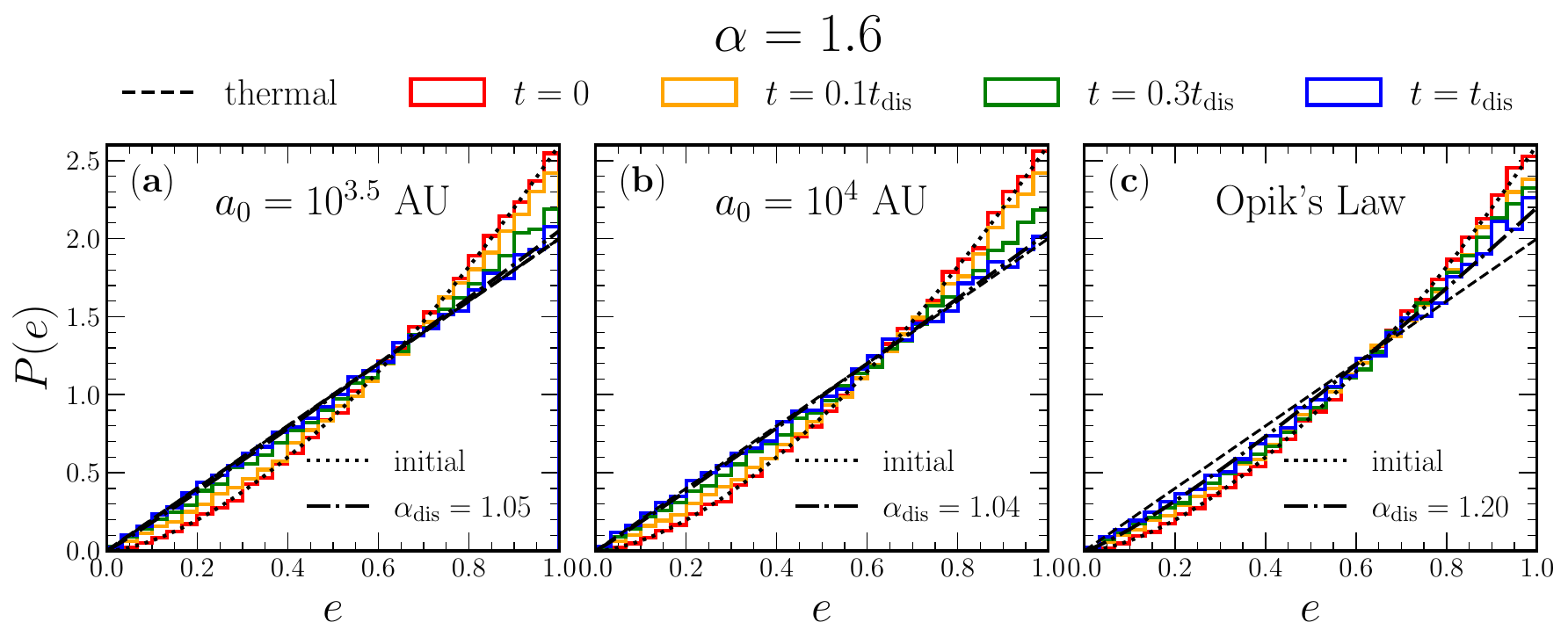}
    \caption{As in Figure \ref{fig:eccentricity_DF_alpha_pt4}, except for an initially superthermal eccentricity DF with $\alpha=1.6$.}
    \label{fig:eccentricity_DF_alpha_1pt6}
\end{figure*}

In Figure \ref{fig:eccentricity_DF_alpha_1pt6} we repeat the same experiment except we change the initial eccentricity DF to be strongly superthermal, with $\alpha = 1.6$. The resulting evolution is almost a mirror image of that in Figure \ref{fig:eccentricity_DF_alpha_pt4}, where this time the DF remains superthermal while being driven towards its final thermal state. This trend is important, since it means that a superthermal DF can be preserved for many Gyr as long as the initial DF is itself sufficiently superthermal.

In general, we found that regardless of the initial DF, $P(e, t)$ gradually `crawls' towards the thermal distribution $P_\mathrm{th}(e)=2e$. The evolution of $P(e)$ due to stellar encounters is less dramatic than the `sloshing' behavior that we observed during the first $\sim t_\mathrm{Gal}$  of evolution under Galactic tides, when phase mixing was the most violent (see Paper I). We emphasize that in none of our numerical solutions of the Fokker-Planck equation \eqref{eq:Fokker_a_esq_simpler} did we ever find an initially subthermal DF becoming superthermal at any time, or vice versa.


\subsubsection{Present-day semimajor axis and eccentricity DFs in a given semimajor axis bin}
\label{sec:Present_Day}


Up to now, in order to calculate the eccentricity DF $P(e)$ we have marginalized over all $a$ values of the full DF $f(a,\esq,t)$. However, if we are to draw astrophysical conclusions and compare to the GAIA data \citep{Hwang21}, we must instead look at the eccentricity DF \textit{within a given final semimajor axis bin}. We should also consider the effect of including a nontrivial birth history, i.e. not simply assume that all binaries were born at $t=0$. That is what we will do in the remainder of this section. Throughout, we again assume that the initial DF is of the form \eqref{eq:DF_separable}, and we always take the semimajor axis-dependent part to be given by Opik's law with $a_\mathrm{min}=400$\,AU.

\begin{figure*}
    \centering
    \includegraphics[width=0.8\textwidth]{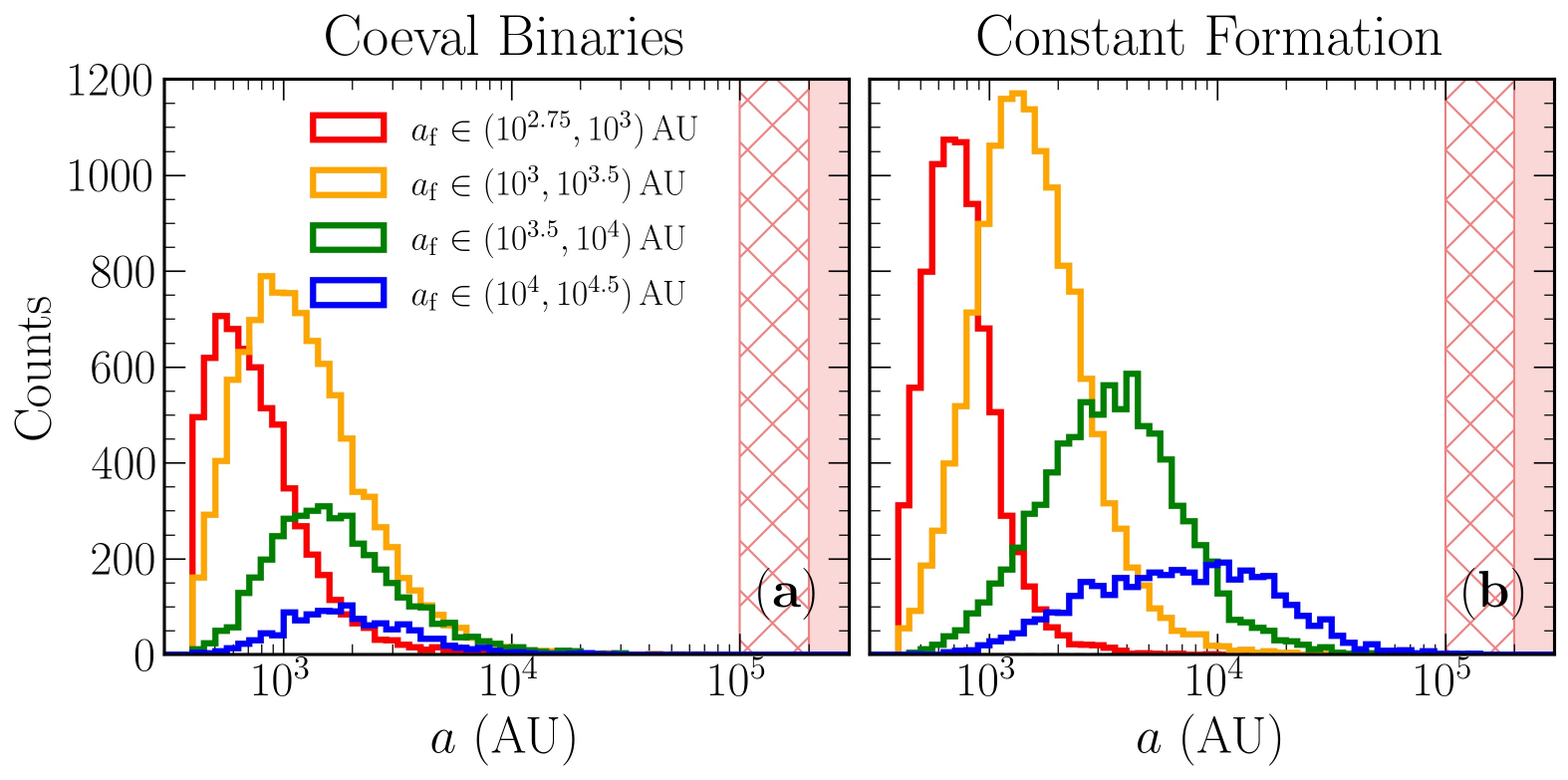}
    \caption{The distribution of \textit{initial} semimajor axis values, for binaries that fall within a given \textit{final} semimajor axis $a_\mathrm{f}$ range (after $10$\,Gyr of evolution). In panel (a), all binaries are born in a single burst at $t=0$,whereas in panel (b) they are assumed to form at a constant rate over $10$\,Gyr. In both cases, the birth  distribution of the binaries is taken to be Opik's law. {The `fringe' region and disruption region are again shown with red hatches and solid red shading respectively.}}
    \label{fig:a_finalbins}
\end{figure*}

To begin with, we consider the solution from panel (c) of Figure \ref{fig:a_histograms}, which shows the evolution of an initially Opik distribution of semimajor axes (we emphasize again that the evolution of the semimajor axis DF is independent of the eccentricity DF). We then ask, given the surviving binaries in a particular semimajor axis bin at $t=10$\,Gyr, what is the distribution of their birth semimajor axes? In Figure \ref{fig:a_finalbins} we plot histograms of these initial semimajor axes with different colored lines for the same four final semimajor axis bins as reported in \cite{Hwang21}'s data\footnote{Strictly speaking, \cite{Hwang21} binned their binaries by separation distance, not semimajor axis, but this difference is not significant at the level of accuracy we are dealing with here.}. Panel (a) shows the results assuming the binaries were all born at $t=0$ (`Coeval Binaries'), whereas panel (b) shows the corresponding result if instead binaries are born at random times $t \in (0, 10\,\mathrm{Gyr})$ (`Constant Formation').

Though neither birth history is particularly realistic, they are extreme cases which provide bounds to the birth history that wide binaries likely underwent in reality. We see that the shape of the distributions look rather similar from panels (a) to (b), but there are many more binaries in total that survive to $t=10$\,Gyr in the constant formation scenario, since younger binaries have had less time to be disrupted. Related to this, we notice that in each panel, even though we allow binaries to form with initial semimajor axis values as large as $2\times 10^5\,$AU, the majority of binaries observed after $10\,$Gyr with $a_\mathrm{f}\in(10^{4}, 10^{4.5})\,$AU (blue histograms) are formed at much smaller $a$.  Of course, this effect is particularly pronounced for the coeval ensemble (panel (a)).  {We also see that almost none of the surviving binaries were born in the fringe $a>10^5$AU, helping to justify our neglect of the complicated fringe dynamics (\S\ref{sec:Approximation_Weak})}. 

\begin{figure*}
    \centering
    \includegraphics[width=0.8\textwidth]{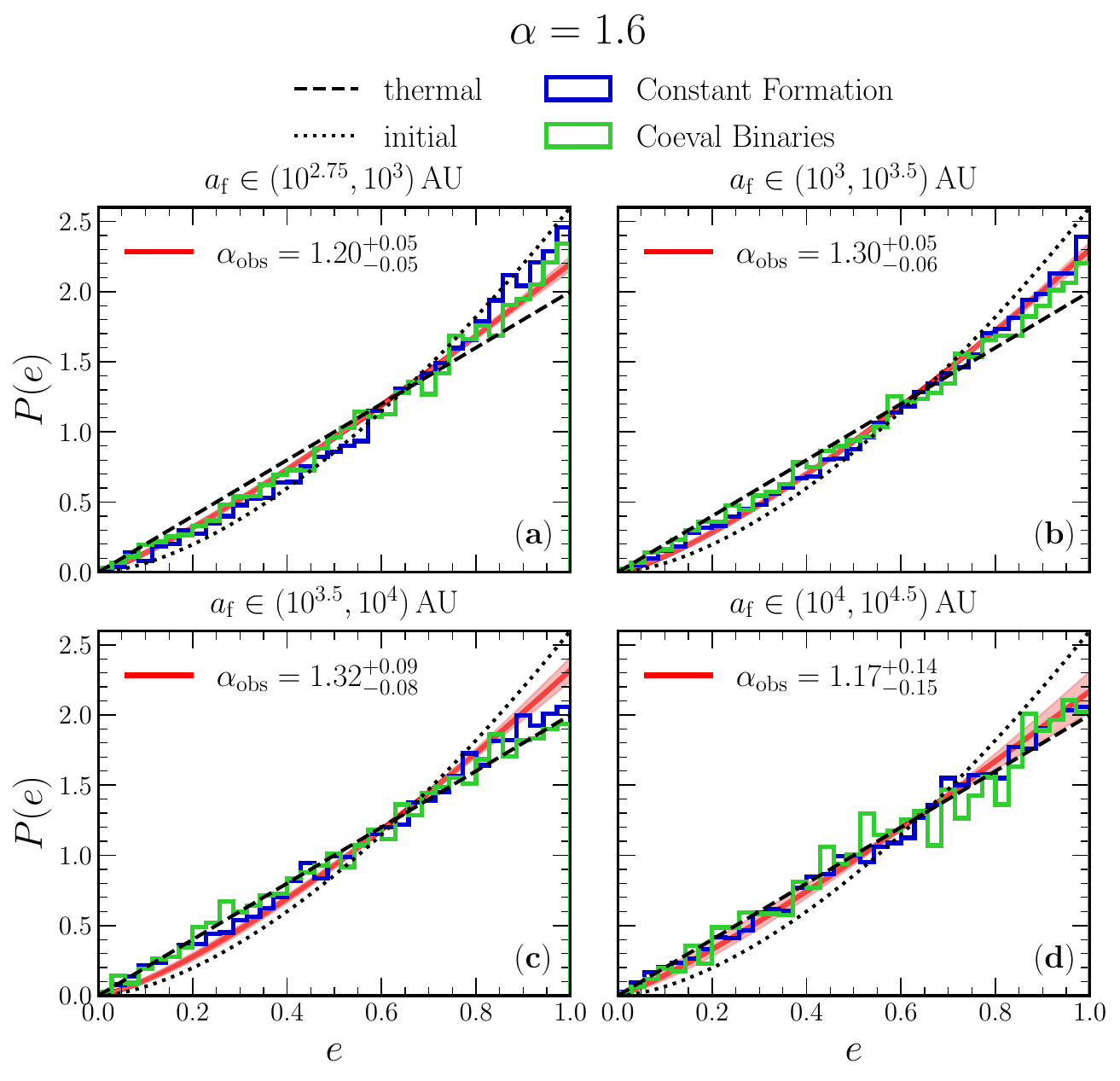}
    \caption{The present-day ($t=10$\,Gyr) eccentricity distribution $P(e)$ for binaries in a given semimajor axis bin. We always assume the binaries were born with Opik's law in semimajor axis and with an initially superthermal eccentricity DF with $\alpha=1.6$. In each panel, we also show in red the power law index inferred from the GAIA data by \citet{Hwang21} for that particular bin, and the shaded region indicates the $\pm1\sigma$ uncertainty of power-law indices about the central value.}
    \label{fig:P_observed_comparison}
\end{figure*}

In Figure \ref{fig:P_observed_comparison} we plot the present-day ($t=10$ Gyr) eccentricity DF $P(e)$ in the same final semimajor axis bins as in Figure \ref{fig:a_finalbins}, again for both coeval and constant-formation birth histories (shown with green and blue histograms respectively). In this case, we chose the eccentricities to be initially distributed according to a superthermal $\alpha = 1.6$ power law, shown with a black dotted line in each panel. We also superpose in red the corresponding power-law eccentricity DF inferred from the GAIA data by \cite{Hwang21}.
We see that in all four panels, the eccentricity DF after $10$ Gyr is still superthermal for the constant formation scenario; this is natural, since new highly eccentric binaries are always being born in this case. Of course the eccentricity DF is \textit{more} superthermal at smaller semimajor axes, due to the longer thermalization timescale (equation \eqref{eq:ionize_numerical}). On the other hand, the coeval formation scenario produces final eccentricity DFs that are essentially indistinguishable from the thermal DF above a critical semimajor axis of $a_\mathrm{f} = 10^{3.5}$AU (panels (c) and (d)), and only weakly superthermal at smaller $a_\mathrm{f}$.

\begin{figure}
    \centering
    \includegraphics[width = 0.45\textwidth]{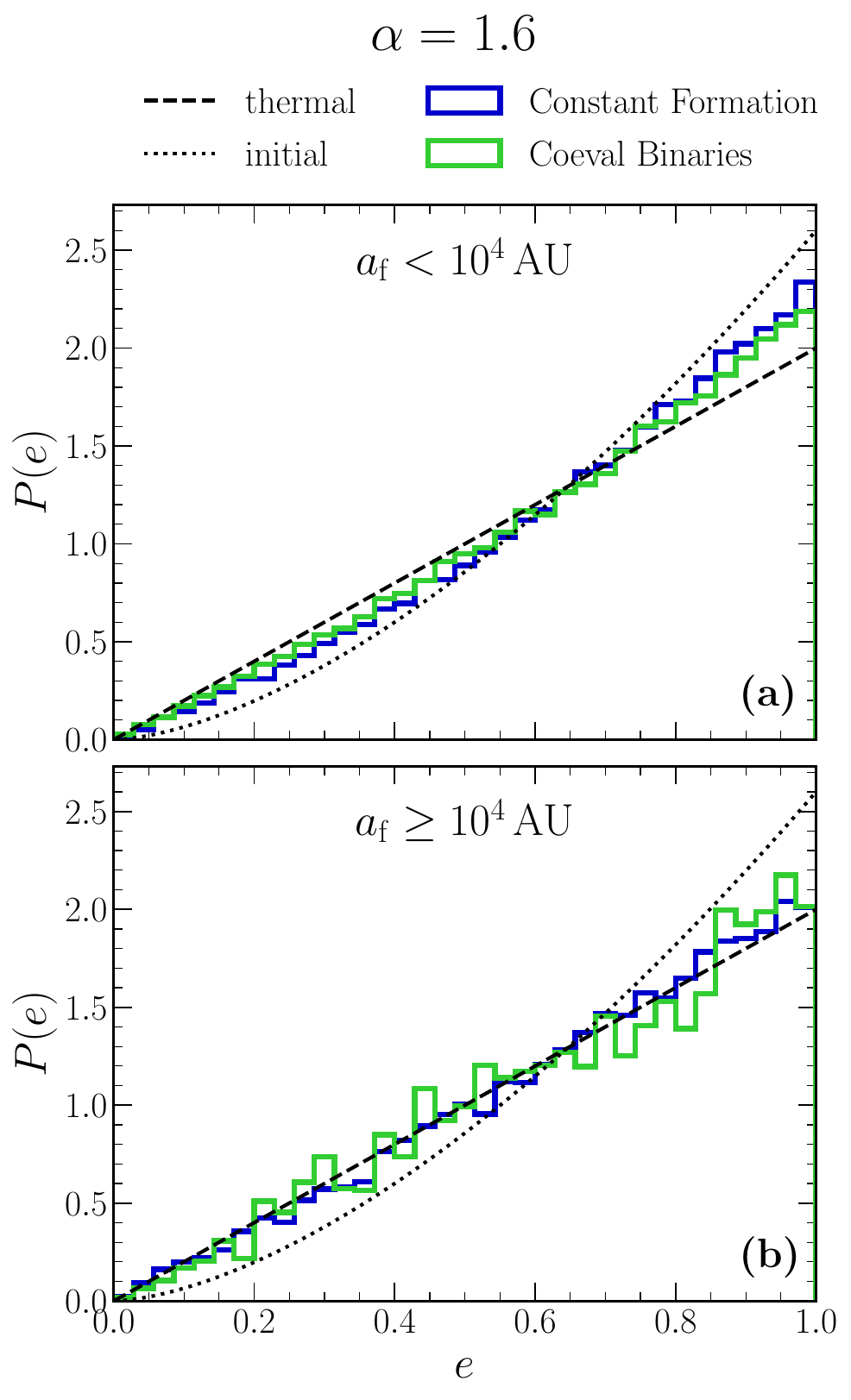}
    \caption{The same data as in Figure \ref{fig:P_observed_comparison}, but binning the binaries according to whether $a_\mathrm{f}$ is below or above $10^4\,$AU.
    {Strictly speaking, panel (b) includes `fringe' binaries with $a>10^5$\,AU, but these are so few in number that excluding them would not noticeably alter the result.}}
    \label{fig:P_overunder1e4}
\end{figure}

To make this semimajor axis and birth history dependence clearer, we present Figure \ref{fig:P_overunder1e4}, which uses exactly the same data as in Figure \ref{fig:P_observed_comparison} except we now use only two bins, splitting the binaries according to whether their $a_\mathrm{f}$ lies below or above $10^4$\,AU. Again, the basic trend is that constant formation scenario and smaller semajor axes give rise to a more superthermal eccentricity DF, whereas a coeval formation scenario and large semimajor axes give rise to a near-thermal DF. {Although in Figure \ref{fig:P_overunder1e4}b we include binaries in the `fringe' with $a > 10^5$\,AU, these make up fewer than $3\%$ of the surviving binaries (see Figure \ref{fig:a_histograms}) and removing them from consideration would not alter our conclusions regardless of their eccentricities distribution.}

Broadly speaking, it is clear from Figure \ref{fig:P_observed_comparison} that, if the only dynamical effect at play was stellar encounters, then a constant formation scenario in which binaries form with an initial $\alpha = 1.6$ eccentricity DF would provide a reasonable fit to the data above $a=10^3$AU. In reality though, since Galactic tides also drive any initial DF closer to thermal, the initial $\alpha$ must actually be even larger than $1.6$ --- see \S\ref{sec:Astrophysical_Implications}.


\section{Discussion}
\label{sec:Discussion}



\subsection{Astrophysical implications and predictions}
\label{sec:Astrophysical_Implications}


In Paper I we showed that under the assumption that binaries are born with random orientations, Galactic tides are incapable of ever transforming an initially subthermal eccentricity DF into the observed superthermal one. {In this paper, again subject to a few weak assumptions (see \S\ref{sec:approximations}), we have seen that this is also true for stellar encounters.
Their role is to} gradually drive any initial eccentricity DF towards thermal on the disruption timescale \eqref{eq:ionize_numerical}.

Strictly speaking, we do not know what the combined impact of Galactic tides and stellar encounters is (and they are both of similar importance for binaries with $a\sim 10^4$\,AU, see Figure \ref{fig:timescales}). Despite this caveat, by combining the results of Paper I with those of the present paper it seems safe to conclude that 
{at least for binaries with $a\in (10^4, 10^{4.5})$ AU today}, \textit{dynamical interactions cannot be responsible for producing superthermal wide binary eccentricity distributions}. In other words, no subset of (initially isotropically oriented) wide binaries whose initial eccentricity DF is subthermal can ever be transformed into a superthermal distribution (as is observed) through dynamical effects.

In fact, we can go further than this. We know that over a lifetime of several Gyr, Galactic tides and stellar encounters will tend to drive any eccentricity DF closer to thermal. It follows that the binaries we observe today must have been born with an \textit{even more superthermal} (higher $\alpha$) initial eccentricity DF than is now measured. Based upon the results of \S\ref{sec:Present_Day} and Paper I, we suspect that whatever the dominant binary formation mechanism, the initial DF it produced must have had an effective power law of at least $\alpha_\mathrm{i} \gtrsim 1.6$. This result places strong constraints on wide binary formation mechanisms, and allows one to rule out several of the channels previously deemed possible. For instance, the formation of wide binaries during the dissolution of young stellar clusters \citep{kouwenhoven2010formation} gives rise to an initially near-thermal eccentricity DF ($\alpha_\mathrm{i}\approx 1)$, and so cannot be the dominant formation channel. The same goes for the formation of wide binaries in stellar streams, which gives a slightly subthermal $\alpha_\mathrm{i}$ \citep{penarrubia2021creation}. The requirement that $\alpha_\mathrm{i}\gtrsim 1.6$ \textit{does} seem to be consistent with the highly simplistic calculations of \citep{xu2023wide} based on a turbulent fragmentation scenario, but more numerical work needs to be done to make their claims reliable.
{A superthermal DF also seems to be a natural outcome of wide binary formation via chaotic three-body encounters \citep{atallah2024binary,ginat2024three}}.

Moreover, the superthermal eccentricity DFs observed by \cite{Hwang21} did not have the same power law $\alpha$ at all $a$. Instead, as reported in Figure \ref{fig:P_observed_comparison}, the value of $\alpha$ was around $1.2$ for $a \sim 10^3$\,AU and $a \gtrsim 10^4$\,AU, with a peak at $\alpha \sim 1.32$ in-between. On the one hand, the decrease in $\alpha$ towards the largest semimajor axes is consistent with what one would naively expect from our picture (see Figure \ref{fig:P_observed_comparison}a):  those are the binaries most susceptible to dynamical interactions, and so are the most susceptible to being driven back towards the thermal distribution. However, such a statement is complicated by the fact that we do not know the birth history of wide binaries nor their initial $a$ distribution, and the final eccentricity DF at each $a$ involves a convolution of these factors (\S\S\ref{sec:Source_Sink} and \ref{sec:Present_Day}). Thus, we can make no robust predictions about the dependence of $\alpha$ on $a$ given the currently available data.

However, there is one robust prediction of our theory which is independent of the unknown birth history and semimajor axis distribution. Namely, $\alpha$ \textit{should be a decreasing function of stellar age}. This prediction is already testable with current data, and will become even more so with future GAIA data releases, since these will include the full radial velocity catalog which will allow better calibration of binary ages based on the age-velocity dispersion relation \citep{mackereth2019dynamical}. It is already known that the wide binary \textit{separation} distribution gets steeper for older binaries \citep{tian2019separation}, but the corresponding measurement for the eccentricity distribution has not yet been made. A measurement of the joint $(a,e)$ distribution as a function of binary age, combined with careful modeling,
might provide very strong constraints on wide binary formation processes.


\subsection{Comparison to previous literature}
\label{sec:lit_comparison}


{The use of statistical methods to follow the evolution and disruption of bound orbits under fluctuating forces was pioneered by  \cite{chandrasekhar1941statistical}.}
The generic Fokker-Planck equation in energy and angular momentum for ensembles of orbits undergoing isotropic, impulsive encounters events is well known (e.g. \citealt{cohn1979numerical,spitzer1987,BT}). However, it has not often been applied to \textit{binaries}, and when it has, usually in terms of semimajor axis only \citep{Weinberg1987,jiang2010evolution}. An exception is the closely related study by \cite{cohn1978stellar}, who used the Fokker-Planck equation to follow the evolution of orbits around a supermassive black hole in the combined energy and angular momentum space. On the more formal side, several useful results were derived by \cite{penarrubia2019stochastic}, including the more general case where the perturbers are extended structures, rather than point masses\footnote{However, one should be careful not to misuse his results. In particular, one must note that $\langle \Delta J \rangle$ and $\langle \Delta E \Delta J \rangle$ are in general  nonzero even under isotropic encounters (Appendix \ref{sec:Encounter_Average}; c.f. \S4.2 of \citealt{penarrubia2019stochastic}).}.  {\cite{penarrubia2019stochastic,penarrubia2021creation} has also shown how to deal with the most weakly-bound binaries, i.e. those in the `fringe' (here meaning $a\gtrsim 10^5$AU), and investigated not only how they are disrupted but also how they can recombine. As discussed in \S\ref{sec:Approximation_Weak}, this regime is outside of the main scope of our paper (and of most observations).}

It is worth mentioning that mathematically, the problem of the orbital evolution of wide binaries is very similar to that of the evolution of long-period Oort comets \citep{Heisler1986-eb}. The analytic tools developed here and in Paper I are therefore ideally suited to study Oort comet dynamics, except when they are at extremely high eccentricities (very small perihelia). (In that limit, the Oort comets can experience interactions with giant planets which are not covered by our formalism.
Their angular momenta can also become so small that stellar encounters can no longer be considered weak in the sense of \S\ref{sec:Approximation_Weak} --- see \cite{Collins2010}).
Furthermore, our results might also be applied to study other exotic objects in the Galactic field like very wide black hole binaries and triples \citep{michaely2019gravitational,michaely2020high} and merging/chaotic binary and triple stars \citep{kaib2014very,grishin2022chaotic}.


\section{Summary}
\label{sec:summary}


In this work, we have studied the effect of stellar encounters on the distribution of semimajor axes $a$ and eccentricities $e$ of wide stellar binaries. Our key findings can be summarized as follows:
\begin{itemize}
    \item The stellar encounters that affect wide binaries are predominantly weak, impulsive and penetrative, and impinge upon each binary in an approximately isotropic fashion. We derived a Fokker-Planck equation in $(a,e^2)$ space which describes the evolution of an ensemble of wide binaries under many such encounters.
    \item Stellar encounters drive the eccentricities of wide binaries towards a thermal distribution at each semimajor axis $a$, while also driving binaries to larger $a$ on average until they are disrupted around $a_\mathrm{dis} \approx 2\times 10^5$\,AU. The simultaneous thermalization and disruption processes happen on the same timescale, which for solar mass binaries is roughly $\sim 4 \, \mathrm{Gyr} \, (a/10^4\,\mathrm{AU})^{-1}$.
    \item Given these results (and those of Paper I), dynamical interactions cannot produce the observed superthermal eccentricity DF of wide binaries with $a \in (10^3, 10^{4.5})$\,AU from any DF that was not already superthermal. In fact, the eccentricity DF of wide binaries at birth must have been significantly more superthermal than is observed today.  
    \item The requirement of an initially very superthermal eccentricity DF places strong constraints upon wide binary formation mechanisms. For instance, it seems that young stellar cluster dissolution cannot be the dominant formation channel.
    \item If our theory is correct, then the $\alpha$ parameter characterizing eccentricity power-law DFs should be a decreasing function of wide binary age. This prediction can be tested with current and upcoming observational data.
\end{itemize}


\section*{Data Availability}


The numerical simulation results used in this paper will be shared on request to the corresponding author.


\section*{Acknowledgements}

We are grateful to the referee, Jorge Pe{\~n}arrubia, for a careful reading and constructive criticism which improved the paper.
We also thank Scott Tremaine and Hsiang-Chih Hwang for detailed comments on the manuscript, Eve Ostriker for a helpful discussion of the role of molecular clouds in this problem, and Mordecai-Mark Mac Low for pointing out important literature.
C.H. is supported by the John N. Bahcall Fellowship Fund at the Institute for Advanced Study. S.M. acknowledges support from the National Science Foundation Graduate Research Fellowship under Grant No. DGE-2039656.

\bibliographystyle{mnras}
\bibliography{bibliography}

\begin{thebibliography}{}
\makeatletter
\relax
\def\mn@urlcharsother{\let\do\@makeother \do\$\do\&\do\#\do\^\do\_\do\%\do\~}
\def\mn@doi{\begingroup\mn@urlcharsother \@ifnextchar [ {\mn@doi@}
  {\mn@doi@[]}}
\def\mn@doi@[#1]#2{\def\@tempa{#1}\ifx\@tempa\@empty \href
  {http://dx.doi.org/#2} {doi:#2}\else \href {http://dx.doi.org/#2} {#1}\fi
  \endgroup}
\def\mn@eprint#1#2{\mn@eprint@#1:#2::\@nil}
\def\mn@eprint@arXiv#1{\href {http://arxiv.org/abs/#1} {{\tt arXiv:#1}}}
\def\mn@eprint@dblp#1{\href {http://dblp.uni-trier.de/rec/bibtex/#1.xml}
  {dblp:#1}}
\def\mn@eprint@#1:#2:#3:#4\@nil{\def\@tempa {#1}\def\@tempb {#2}\def\@tempc
  {#3}\ifx \@tempc \@empty \let \@tempc \@tempb \let \@tempb \@tempa \fi \ifx
  \@tempb \@empty \def\@tempb {arXiv}\fi \@ifundefined
  {mn@eprint@\@tempb}{\@tempb:\@tempc}{\expandafter \expandafter \csname
  mn@eprint@\@tempb\endcsname \expandafter{\@tempc}}}

\bibitem[\protect\citeauthoryear{Atallah, Weatherford, Trani  \& Rasio}{Atallah
  et~al.}{2024}]{atallah2024binary}
Atallah D.,  Weatherford N.~C.,  Trani A.~A.,   Rasio F.,  2024, arXiv preprint
  arXiv:2402.12429

\bibitem[\protect\citeauthoryear{Bahcall, Hut  \& Tremaine}{Bahcall
  et~al.}{1985}]{bahcall1985maximum}
Bahcall J.,  Hut P.,   Tremaine S.,  1985, The Astrophysical Journal, 290, 15

\bibitem[\protect\citeauthoryear{Bar-Or, Kupi  \& Alexander}{Bar-Or
  et~al.}{2013}]{bar2013stellar}
Bar-Or B.,  Kupi G.,   Alexander T.,  2013, The Astrophysical Journal, 764, 52

\bibitem[\protect\citeauthoryear{{Bate}}{{Bate}}{2014}]{Bate2014}
{Bate} M.~R.,  2014, Monthly Notices of the Royal Astronomical Society, 442,
  285

\bibitem[\protect\citeauthoryear{{Binney} \& {Tremaine}}{{Binney} \&
  {Tremaine}}{2008}]{BT}
{Binney} J.,  {Tremaine} S.,  2008, {Galactic Dynamics: Second Edition}.
Princeton University Press

\bibitem[\protect\citeauthoryear{Chandrasekhar}{Chandrasekhar}{1941}]{chandrasekhar1941statistical}
Chandrasekhar S.,  1941, Astrophysical Journal, vol. 94, p. 511, 94, 511

\bibitem[\protect\citeauthoryear{Cohn}{Cohn}{1979}]{cohn1979numerical}
Cohn H.,  1979, The Astrophysical Journal, 234, 1036

\bibitem[\protect\citeauthoryear{Cohn \& Kulsrud}{Cohn \&
  Kulsrud}{1978}]{cohn1978stellar}
Cohn H.,  Kulsrud R.,  1978, The Astrophysical Journal, 226, 1087

\bibitem[\protect\citeauthoryear{Collins \& Sari}{Collins \&
  Sari}{2010}]{Collins2010}
Collins B.~F.,  Sari R.,  2010, The Astronomical Journal, 140, 1306

\bibitem[\protect\citeauthoryear{Cournoyer-Cloutier et~al.,}{Cournoyer-Cloutier
  et~al.}{2021}]{cournoyer2021implementing}
Cournoyer-Cloutier C.,  et~al., 2021, Monthly Notices of the Royal Astronomical
  Society, 501, 4464

\bibitem[\protect\citeauthoryear{Ginat \& Perets}{Ginat \&
  Perets}{2024}]{ginat2024three}
Ginat Y.~B.,  Perets H.~B.,  2024, Monthly Notices of the Royal Astronomical
  Society, 531, 739

\bibitem[\protect\citeauthoryear{Grishin \& Perets}{Grishin \&
  Perets}{2022}]{grishin2022chaotic}
Grishin E.,  Perets H.~B.,  2022, Monthly Notices of the Royal Astronomical
  Society, 512, 4993

\bibitem[\protect\citeauthoryear{{Hamilton}}{{Hamilton}}{2022}]{Hamilton22}
{Hamilton} C.,  2022, The Astrophysical Journal Letters, 929, L29

\bibitem[\protect\citeauthoryear{Heggie}{Heggie}{1975}]{heggie1975binary}
Heggie D.~C.,  1975, Monthly Notices of the Royal Astronomical Society, 173,
  729

\bibitem[\protect\citeauthoryear{Heisler \& Tremaine}{Heisler \&
  Tremaine}{1986}]{Heisler1986-eb}
Heisler J.,  Tremaine S.,  1986, Icarus, 65, 13

\bibitem[\protect\citeauthoryear{{Hwang}, {Ting}  \& {Zakamska}}{{Hwang}
  et~al.}{2022a}]{Hwang21}
{Hwang} H.-C.,  {Ting} Y.-S.,   {Zakamska} N.~L.,  2022a, Monthly Notices of
  the Royal Astronomical Society, 512, 3383

\bibitem[\protect\citeauthoryear{Hwang et~al.,}{Hwang
  et~al.}{2022b}]{hwang2022widethick}
Hwang H.-C.,  et~al., 2022b, Monthly Notices of the Royal Astronomical Society,
  513, 754

\bibitem[\protect\citeauthoryear{Jiang \& Tremaine}{Jiang \&
  Tremaine}{2010}]{jiang2010evolution}
Jiang Y.-F.,  Tremaine S.,  2010, Monthly Notices of the Royal Astronomical
  Society, 401, 977

\bibitem[\protect\citeauthoryear{Kaib \& Raymond}{Kaib \&
  Raymond}{2014}]{kaib2014very}
Kaib N.~A.,  Raymond S.~N.,  2014, The Astrophysical Journal, 782, 60

\bibitem[\protect\citeauthoryear{Kouwenhoven, Goodwin, Parker, Davies, Malmberg
   \& Kroupa}{Kouwenhoven et~al.}{2010}]{kouwenhoven2010formation}
Kouwenhoven M.,  Goodwin S.,  Parker R.~J.,  Davies M.~B.,  Malmberg D.,
  Kroupa P.,  2010, Monthly Notices of the Royal Astronomical Society, 404,
  1835

\bibitem[\protect\citeauthoryear{Mackereth et~al.,}{Mackereth
  et~al.}{2019}]{mackereth2019dynamical}
Mackereth J.~T.,  et~al., 2019, Monthly Notices of the Royal Astronomical
  Society, 489, 176

\bibitem[\protect\citeauthoryear{Magorrian \& Tremaine}{Magorrian \&
  Tremaine}{1999}]{magorrian1999rates}
Magorrian J.,  Tremaine S.,  1999, Monthly Notices of the Royal Astronomical
  Society, 309, 447

\bibitem[\protect\citeauthoryear{Merritt}{Merritt}{2015}]{merritt2015gravitational}
Merritt D.,  2015, The Astrophysical Journal, 804, 52

\bibitem[\protect\citeauthoryear{Michaely \& Perets}{Michaely \&
  Perets}{2019}]{michaely2019gravitational}
Michaely E.,  Perets H.~B.,  2019, The Astrophysical Journal Letters, 887, L36

\bibitem[\protect\citeauthoryear{Michaely \& Perets}{Michaely \&
  Perets}{2020}]{michaely2020high}
Michaely E.,  Perets H.~B.,  2020, Monthly Notices of the Royal Astronomical
  Society, 498, 4924

\bibitem[\protect\citeauthoryear{{Modak} \& {Hamilton}}{{Modak} \&
  {Hamilton}}{2023}]{MH23}
{Modak} S.,  {Hamilton} C.,  2023, Monthly Notices of the Royal Astronomical
  Society, 524, 3102

\bibitem[\protect\citeauthoryear{Pe{\~n}arrubia}{Pe{\~n}arrubia}{2019}]{penarrubia2019stochastic}
Pe{\~n}arrubia J.,  2019, Monthly Notices of the Royal Astronomical Society,
  484, 5409

\bibitem[\protect\citeauthoryear{Pe{\~n}arrubia}{Pe{\~n}arrubia}{2021}]{penarrubia2021creation}
Pe{\~n}arrubia J.,  2021, Monthly Notices of the Royal Astronomical Society,
  501, 3670

\bibitem[\protect\citeauthoryear{Reipurth \& Mikkola}{Reipurth \&
  Mikkola}{2012}]{reipurth2012formation}
Reipurth B.,  Mikkola S.,  2012, Nature, 492, 221

\bibitem[\protect\citeauthoryear{Rozner \& Perets}{Rozner \&
  Perets}{2023}]{rozner2023born}
Rozner M.,  Perets H.~B.,  2023, arXiv preprint arXiv:2304.02029

\bibitem[\protect\citeauthoryear{{Spitzer}}{{Spitzer}}{1987}]{spitzer1987}
{Spitzer} L.,  1987, {Dynamical evolution of globular clusters}.
Princeton University Press

\bibitem[\protect\citeauthoryear{Stegmann et~al.,}{Stegmann
  et~al.}{2024}]{stegmann2024close}
Stegmann J.,  et~al., 2024, arXiv preprint arXiv:2405.02912

\bibitem[\protect\citeauthoryear{Sun et~al.,}{Sun
  et~al.}{2022}]{sun2022molecular}
Sun J.,  et~al., 2022, The Astronomical Journal, 164, 43

\bibitem[\protect\citeauthoryear{Tian, El-Badry, Rix  \& Gould}{Tian
  et~al.}{2019}]{tian2019separation}
Tian H.-J.,  El-Badry K.,  Rix H.-W.,   Gould A.,  2019, The Astrophysical
  Journal Supplement Series, 246, 4

\bibitem[\protect\citeauthoryear{Tokovinin}{Tokovinin}{2020}]{Tokovinin2020-go}
Tokovinin A.,  2020, Monthly Notices of the Royal Astronomical Society, 496,
  987

\bibitem[\protect\citeauthoryear{{Vom Scheidt}}{{Vom
  Scheidt}}{1994}]{Numerical_SDEs}
{Vom Scheidt} J.,  1994, Zeitschrift Angewandte Mathematik und Mechanik, 74,
  332

\bibitem[\protect\citeauthoryear{Weinberg, Shapiro  \& Wasserman}{Weinberg
  et~al.}{1987}]{Weinberg1987}
Weinberg M.~D.,  Shapiro S.~L.,   Wasserman I.,  1987, The Astrophysical
  Journal, 312, 367

\bibitem[\protect\citeauthoryear{Widmark}{Widmark}{2019}]{Widmark2019-wp}
Widmark A.,  2019, Astronomy \& Astrophysics. Supplement series, 623, A30

\bibitem[\protect\citeauthoryear{Xu, Hwang, Hamilton  \& Lai}{Xu
  et~al.}{2023}]{xu2023wide}
Xu S.,  Hwang H.-C.,  Hamilton C.,   Lai D.,  2023, The Astrophysical Journal
  Letters, 949, L28

\makeatother
\end{thebibliography}



\appendix


\section{Calculation of the Fokker-Planck coefficients}
\label{sec:FP_EJ}


Consider a distribution function which, by Jeans' theorem and isotropy, depends only on $E$ and $J$. Call this DF $N(E,J,t)$, such that the number of binaries near $(E,J)$ at time $t$ is proportional to $N(E, J, t) \, \md E \md J$. The orbit-averaged Fokker-Planck equation governing the evolution of $N$ is (e.g. \citealt{cohn1979numerical,spitzer1987,merritt2015gravitational}):
\begin{align}
    \pder{N}{t} =& 
    -\pder{}{E} [\overline{\langle{\Delta E}\rangle} N]
    -\pder{}{J} [\overline{\langle{\Delta J}\rangle} N] \nn \\
&    + \frac{1}{2}\pder{^2}{E^2} [ \orbitmeansquareDelta{E} N]
    + \frac{1}{2}\pder{^2}{J^2} [ \orbitmeansquareDelta{J} N] \nn \\
&        + \frac{\partial^2 }{\partial E \partial J} [ 
        \overline{\langle{\Delta E \Delta J}\rangle} N].
        \label{eq:Fokker_E_J}
\end{align}
The brackets here refer to averages over a large number of kicks at a fixed orbital position $\bm{r}$ and velocity $\bm{v}$, and the overline refers to a subsequent average over the Keplerian orbit.

To derive expressions for the drift and diffusion coefficients, we begin with the fact that impulsive encounters alter only the binary's relative velocity vector, $\bm{v} \to \bm{v} + \Delta \bm{v}$, without changing its separation vector $\bm{r}$; the changes to $E$ and $\bm{J}$ are then
\begin{equation}
    \label{eq:DeltaE_DeltaJ}
    \Delta E = \bm{v} \cdot \Delta \bm{v} + \frac{1}{2}(\Delta \bm{v})^2, \,\,\,\,\,\,\,\,\,\, \Delta \bm{J} = \bm{r} \times \Delta \bm{v}.
\end{equation}


\subsection{Averaging over encounters}
\label{sec:Encounter_Average}


For isotropic perturbers, $\Delta \bm{v}$ is in a random direction, so without any other approximations, an average over such encounters gives:
\begin{equation}
    \meanDelta{E} = \frac{1}{2} \meansquareDelta{\bm{v}}, \,\,\,\,\,\,\,\,\,\, \meanDelta{\bm{J}} = 0.
\end{equation}
Next, squaring $\Delta E$ from \eqref{eq:DeltaE_DeltaJ}, keeping only terms up to second order in small quantities, and averaging over isotropic encounters,  we find
\begin{equation}
   \langle (\Delta E)^2 \rangle \approx \frac{1}{3} v^2 \langle (\Delta \bm{v})^2\rangle.
\end{equation}
Similarly, we put
\begin{equation}
    \label{eq:Delta_J_generic}
    \Delta J = \vert \bm{J} + \Delta \bm{J} \vert - J.
\end{equation}
Expanding this for small $\Delta \bm{J}$, keeping only the terms up to second order, and encounter-averaging, we arrive at
\begin{equation}
   \langle  \Delta J \rangle \approx \frac{\langle (\Delta \bm{J})^2 \rangle }{4J} > 0.
\end{equation}
Likewise,
\begin{equation}
   \langle  (\Delta J)^2 \rangle \approx \frac{\langle (\Delta \bm{J})^2 \rangle }{2}.
\end{equation}
In each of the last two cases we need to evaluate $\meansquareDelta{\bm{J}}$. This is easy: with no approximation other than isotropy we have from \eqref{eq:DeltaE_DeltaJ} that
\begin{equation}
    \meansquareDelta{\bm{J}} = \frac{2}{3}  r^2 \langle (\Delta \bm{v})^2\rangle.
\end{equation}
Finally, we require $\langle \Delta E \Delta J \rangle$. Multiplying equations  \eqref{eq:DeltaE_DeltaJ} and \eqref{eq:Delta_J_generic}, expanding up to second order in small quantities, and averaging over isotropic encounters, we arrive after some algebra at 
\begin{equation}
    \langle  \Delta E \Delta J \rangle =  \frac{1}{3} J  \langle (\Delta \bm{v} )^2 \rangle.
\end{equation}

In the main text we use not $(E, J)$, but rather the more natural coordinates $(a,\varepsilon)$ where $\esq \equiv e^2$. Then the Fokker-Planck equation \eqref{eq:Fokker_E_J} turns into equation \eqref{eq:Fokker_a_esq_simpler}, where $f$ is the DF in this new space. Using equations \eqref{eq:energy} and \eqref{eq:J_ae}, it is easy to convert the $(E, J)$ drift and diffusion coefficients from above into $(a, \esq)$ coefficients:
\begin{align}
    \label{eq:a_Drift}
    \meanDelta{a} & = \frac{Gm_\mathrm{b}}{2E^2} \meanDelta{E} - \frac{Gm_\mathrm{b}}{2E^3}\meansquareDelta{E} \nn \\
    & = \frac{ Gm_\mathrm{b} \meansquareDelta{\bm{v}}}{4E^2} \left( 1 - \frac{2v^2}{3E}\right),
\\
    \meansquareDelta{a} & = \frac{(Gm_\mathrm{b})^2}{4E^4}\meansquareDelta{E} \nn \\
    \label{eq:a_Diffusion}
    & = \frac{(Gm_\mathrm{b})^2 \meansquareDelta{\bm{v}} v^2 }{12 E^4},
\\
    \meanDelta{\esq} &= \frac{2J^2}{(Gm_\mathrm{b})^2} \meanDelta{E} + \frac{4EJ}{(Gm_\mathrm{b})^2} \meanDelta{J} + \frac{2E}{(Gm_\mathrm{b})^2} \meansquareDelta{J} \nn \\ &\,\,\,\,\,\,+ \frac{4J}{(Gm_\mathrm{b})^2} \langle \Delta E \Delta J \rangle \nn \\ 
    \label{eq:esq_Drift}
    & = \frac{\meansquareDelta{\bm{v}}}{3(Gm_\mathrm{b})^2}(7J^2 + 4E r^2),
\\
    \meansquareDelta{\esq} & = \frac{4J^4}{(Gm_\mathrm{b})^4} \meansquareDelta{E}   + \frac{16E^2J^2}{(Gm_\mathrm{b})^4} \meansquareDelta{J} + \frac{16E J^3}{(Gm_\mathrm{b})^4} \langle \Delta E\Delta J \rangle  \nn \\
    \label{eq:esq_Diffusion}
    & = \frac{4 \meansquareDelta{\bm{v}}}{3(Gm_\mathrm{b})^4}(J^4 v^2 + 4E^2J^2r^2 + 4EJ^4),
\\
   \langle \Delta a \Delta \esq \rangle &= \frac{J^2}{Gm_\mathrm{b} E^2} \langle (\Delta E)^2 \rangle + \frac{2J}{Gm_\mathrm{b} E} \langle \Delta E \Delta J \rangle \nn \\
   & = \frac{\meansquareDelta{\bm{v}} J^2}{3E^2}(v^2+2E).
\end{align}


\subsection{Averaging over orbits}
\label{sec:orbavg}


We may finally transform the encounter-averaged coefficients into orbit-averaged coefficients using the identities (e.g. \citealt{penarrubia2019stochastic}, Appendix C):
\begin{align}
    & \overline{r^2} = a^2\left( 1+\frac{3e^2}{2} \right) = \frac{(Gm_\mathrm{b})^2}{8E^2}(5-3j^2), \\
     & \overline{v^2} = \frac{Gm_\mathrm{b}}{a} = -2E,
\end{align}
and the fact that $\langle (\Delta \bm{v} )^2 \rangle$ does not depend (in the penetrative regime) on the orbital phase. The resulting orbit-averaged coefficients are listed in \S\ref{sec:FP_in_a_esq}.


\section{Proof that $C$ is monotonically decreasing}
\label{sec:Casimir_proof}


Here we  prove that $C$ (equation \eqref{eq:Casimir}) is a monotonically decreasing function of time (equation \eqref{eq:dCdt_negative}), with equality only when $\delta f=0$ everywhere. To do this, let us first multiply \eqref{eq:fluctuation_evolution} by $2 \, \delta f$ and integrate over $a$ and $\esq$. The resulting expression consists of two terms with obvious definitions:
\begin{equation}
    \frac{\md C}{\md t} = \dot{C}_a + \dot{C}_\esq.
\end{equation}
Our aim is then to show that both terms on the right hand side are negative, except if $\delta f =0$ at all $a, \esq$, in which case they are both zero.

Let us deal with the $\dot{C}_\esq$ term first. This is 
\begin{align}
    \dot{C}_\esq & = \int \md a \int_0^1 \md \esq \, \delta f \frac{\partial }{\partial \varepsilon} \left[ D_\varepsilon \frac{\partial \delta f}{\partial \varepsilon}\right] \nn \\
    \label{eq:Cdot-esq}
    & = \int \md a \left\{ \left[D_\esq \delta f \frac{\partial  \delta f}{\partial \esq}\right]_{\esq=0}^{\esq=1} - \int_0^1 \md \esq D_\esq \left( \frac{\partial  \delta f}{\partial \esq} \right)^2 \right\},
\end{align}
where to get the second line we integrated by parts over $\esq$. The boundary term (first term in curly brackets) is zero provided $D_\esq=0$ at $\esq=0$ and $\esq = 1$\footnote{We are also assuming here that $\delta f$ is sufficiently well-behaved at $\esq=0$ and $\esq=1$.}. This must be true for any physically sensible diffusion coefficient, including the one we derived in equation \eqref{eq:epsilon_diff_avg}. The remaining term in \eqref{eq:Cdot-esq} is clearly negative for any $\delta f$ \textit{except} if $\partial \delta f/\partial \esq = 0$ for all $a,\esq$; but as per the definition of $\delta f$, this is only the case if $\delta f=0$ for all $a,\esq$.

We now turn to the $\dot{C}_a$ term, namely 
\begin{equation}
    \dot{C}_a = -2  \int_0^1 \md \esq \int \md a  \, \delta f \frac{\partial }{\partial a } \left[ A_a \delta f - \frac{1}{2}\frac{\partial }{\partial a} (D_a \delta f)\right].
\end{equation}
We integrate this by parts over $a$ from $a_\mathrm{min}$ to $a_\mathrm{max} = a_\mathrm{dis}$. The boundary terms vanish\footnote{The vanishing of the boundary term at $a_\mathrm{max} = a_\mathrm{dis}$ follows immediately from the condition \eqref{eq:disrupting_boundary}. As for the boundary term at $a=a_\mathrm{min}, 
$ this obviously must vanish if we take $a_\mathrm{min}=0$.
However, $a=0$ is well outside the range in which our assumption of penetrative, impulsive encounters is valid.  In practice our numerical simulations start with no binaries below $a_\mathrm{min}=400$ AU, and since the evolution at such small semimajor axes is extremely slow (Figure \ref{fig:timescales}), and these binaries tend to get wider anyway, we find that almost no binaries ever stray below $a_\mathrm{min}$. Hence, we are justified in discarding this boundary term also.} and we are left with
\begin{align}
    \dot{C}_a &=  - \int \md \esq \int \md a D_a \left( \frac{\partial \delta f}{\partial a} \right)^2
    \nn
    \\
    &\,\,\,\,\,\,\,\,\,\,\,\,\,\,\,
    + \int_0^1 \md \esq \int \md a  \, \frac{\partial (\delta f)^2 }{\partial a } \left[ A_a - \frac{1}{2}\frac{\partial D_a}{\partial a}\right].
    \label{eq:Cdota_intermediate}
\end{align}
The first term here is obviously negative or zero.  As for the second term, we again integrate it by parts and throw away boundary terms. For finite $\delta f$, the result is guaranteed to be negative provided the $a$-derivative of the square bracket is positive at all $a$. In particular, if we substitute the expressions \eqref{eq:a_Drift}, \eqref{eq:a_Diffusion}, and \eqref{eq:Jiang_Delta_V_Squared}, then the second line in \eqref{eq:Cdota_intermediate} becomes
\begin{equation}
    - \int_0^1 \md \esq\int \md a \, 2 a\,  (\delta f )^2 \frac{\langle (\Delta \bm{v})^2 \rangle}{3Gm_\mathrm{b}} \left[ 1 - \frac{3}{2\ln \Lambda}\right],
\end{equation}
which is $\leq 0$ provided $\ln \Lambda \geq 3/2$ (which is always true in the impulsive regime, see equation \eqref{eq:Coulomb}). Again, equality holds only if $\delta f=0$ everywhere.

\bsp
\label{lastpage}
\end{document}